\documentclass[11pt,a4paper]{article}

\usepackage[margin=2.5cm]{geometry}
\usepackage{mathtools,amsmath,amssymb,amsfonts,amsthm,makeidx,graphicx}
\usepackage{xcolor}
\usepackage[colorlinks=true,linkcolor=blue,citecolor=blue,urlcolor=blue]{hyperref}
\usepackage{units}
\usepackage{tikz}
\usetikzlibrary{calc}
\usepackage{enumitem}

\newif\ifrevisionmarks
\revisionmarksfalse
\ifrevisionmarks
\newcommand{\rev}[1]{{\color{blue}#1}}
\else
\newcommand{\rev}[1]{#1}
\fi

\def\be{\begin{equation}}
\def\ee{\end{equation}}
\def\ba{\begin{eqnarray}}
\def\ea{\end{eqnarray}}

\def\m{\mu}
\def\n{\nu}
\def\p{\partial}
\def\r{\right}
\def\f{\frac}

\renewcommand{\l}{\left(}

\renewcommand{\O}{\Omega}

\newcommand{\nuMSM}{\ifmmode\nu\mathrm{MSM~}\else$\nu$MSM~\fi}

\usepackage{xifthen}
\newcommand*\diff{\mathrm{d}} 
\newcommand*\ldiff[2][]{ \ifthenelse{\isempty{#1}}{ \diff #2}{\diff^#1#2} \,} 

\begin{document}

\begin{center}

{\LARGE\bfseries The Higgs boson, naturalness, and the Universe\par}

\vspace{10mm}

{\large Mikhail Shaposhnikov}\\[5mm]
{\itshape Institute of Physics, \'Ecole Polytechnique F\'ed\'erale de Lausanne (EPFL),\\ CH-1015 Lausanne, Switzerland}\\[3mm]

\vspace{10mm}

{\bfseries Abstract}
\end{center}

\noindent
The Higgs boson -- the only known fundamental scalar particle -- may have profound implications for both particle physics and cosmology. This review examines how the Higgs field may have influenced the Universe from its earliest moments to its present state. We first discuss particle-physics puzzles, including the naturalness and hierarchy problems, as well as the possible metastability of the electroweak vacuum. We then turn to cosmological scenarios where the Higgs boson could be a driving force, including Higgs inflation, reheating and the onset of the hot Big Bang, the electroweak phase transition, and baryogenesis via electroweak processes.

\section{Introduction}
\label{intro}
There are numerous reviews and textbooks which describe the physics of the Standard Model  (SM) and the role of the Higgs boson in particle physics, see e.g. \cite{ParticleDataGroup:2024cfk}. It is common knowledge that the Higgs boson ``breaks'' the SU(2)xU(1) symmetry down to U(1), it  ``gives'' masses to quarks, charged leptons, and to the intermediate vector bosons $W$ and $Z$. The Yukawa interactions of the Higgs boson to quarks lead to flavour-changing transitions in the weak interactions and induce CP-violation due to the Kobayashi-Maskawa complex phase. 

The SM stands very strongly.  Theoretically, it is described by a renormalisable Quantum Field Theory (QFT)  and does not have any intrinsic inconsistencies (such as Landau poles in the scalar self-coupling and U(1) gauge coupling \cite{Maiani:1977cg, Cabibbo:1979ay, Lindner:1985uk})  up to a very high energy scale, exceeding that of quantum gravity $M_P \sim 10^{19}$ GeV.  Empirically, no deviations from the Standard Model have been observed in any experiment, except for neutrino oscillations (remember, in the canonical SM \cite{Glashow:1961tr, Weinberg:1967tq, Salam:1968rm} neutrinos are exactly massless, leading to the conservation of the e, $\mu$, and $\tau$ leptonic numbers). The simplest resolution of this drawback of the SM, lying entirely within the SM philosophy (same gauge group, same structure of fermionic generations, same Higgs sector), is adding to the SM  three right-handed singlet fermions - ``sterile''  or Majorana neutrinos, making the leptons' left-right structure similar to that of quarks \cite{Minkowski:1977sc, Yanagida:1979as, Gell-Mann:1979vob, Glashow:1979nm, Mohapatra:1979ia}.  This theory will be called  \nuMSM -- Neutrino Minimal Standard Model   \cite{Asaka:2005an} -- in what follows.

Given the success of the SM in particle physics, it is only natural to seek to understand the evolution of the Universe from the very first moments (such as the inflationary era) to the present time in the framework of the SM (or $\nu$MSM, to account for neutrino oscillations) and gravity. Can we describe everything we observe? What is the role of the Higgs field in cosmology?

This review is devoted to possible answers to these questions. We will concentrate mainly on motivations and qualitative ideas, rather than technical details, which can be found in the original literature. It is organised as follows. Section \ref{SM} is a short reminder of the SM and the $\nu$MSM, fixing the notations to be used throughout the paper. In Section \ref{consistency}, we discuss the self-consistency of the Standard Model up to the Planck scale, consider the effective potential at large values of the Higgs field, and discuss a possible metastability of the electroweak vacuum. Section  \ref{Naturalness}  contains a discussion of the hierarchy problem and ``naturalness'' as a criterion to search for new physics beyond the SM (BSM), both associated with the elementary scalar field - the Higgs boson. In Section \ref{EWPT}, we go to the electroweak epoch of the Universe expansion and discuss the symmetry restoration at high temperatures and possibilities of phase transitions in the early Universe, with applications to the SM. In Section \ref{BAU}, we discuss the structure of CP violation in the interactions of the Higgs boson with fermions and outline possibilities for baryogenesis in the SM and $\nu$MSM. We shortly describe a sterile neutrino candidate for Dark Matter, produced because of Yukawa interactions with the Higgs boson, in Section \ref{DM}. In Section \ref{HI},  we review Higgs inflation, with the Higgs boson making the Universe nearly flat, homogeneous, and isotropic, and producing the fluctuations leading to the formation of structures in the Universe.  In addition to inflating the Universe, the Higgs boson serves as the origin of the Hot Big Bang. Section \ref{Conclusions}  presents our conclusions, where we summarise the main points of the paper.

\section{Setting the scene: the Standard Model and the  $\nu$MSM}
\label{SM}
\subsection{The SM}
\label{SM1}
The SM of particle interactions \cite{Glashow:1961tr,Weinberg:1967tq,Salam:1968rm} is based on SU(3)$\times$SU(2)$\times$U(1) gauge symmetry. Besides the gauge bosons, it contains three generations of fermions -  left and right-handed quarks and leptons, and the Higgs field $H$ - a doublet of the SU(2). The theory is renormalisable and is described by a Lagrangian 
\begin{align}
\label{LSM}
\mathcal{L}_{\rm SM} &=
-\frac14 F^a_{\mu\nu}F^{a\mu\nu}  + \bar\psi\, i\gamma^\mu D_\mu \psi
+ (D_\mu H)^\dagger (D^\mu H)
- V(H) + {\cal L}_Y + \text{h.c.}\,.
\end{align}
In the kinetic term, $\psi$ is the generic notation of fermions, and $F^a_{\mu\nu}$  combines all the gauge fields. The Higgs potential is given by
\be
\label{pot}
V(H) = -m^2 H^\dagger H + \lambda (H^\dagger H)^2\,,
\ee
and the covariant derivatives are 
\be
D_\mu = \partial_\mu
- i g_s T^a G^a_\mu
- i g \frac{\tau^i}{2} W^i_\mu
- i g' Y B_\mu\,,
\ee
where $G^a_\mu, W^i_\mu$ and $B_\mu$ are the SU(3), SU(2) and U(1) gauge fields coming with the corresponding group generators and coupling constants $g_s,~g$ and $g'$. The gauge field strengths are 
\be
F^a_{\mu\nu} = \partial_\mu A^a_\nu - \partial_\nu A^a_\mu + g_s f^{abc} A^b_\mu A^c_\nu \,,
\ee
with the structure constants corresponding to the SM group. The Yukawa interaction is 
\be
{\cal L}_Y = \frac{g_W}{\sqrt{2}M_W} \{\bar{Q}_L^\alpha K^{\alpha\beta} M_d^{\beta\beta}
D_R^\beta H +
\bar{Q}_L^\alpha M_u^{\alpha\alpha} U_R^\alpha\tilde{H} + \bar{L}^\alpha M_e^{\alpha\alpha} e_R^\alpha H\},
\label{Yukawa}
\ee
where  $Q_L^\alpha$ and $L^\alpha$ denote the left-handed quark and lepton doublets ($\alpha$ is the generation index), $U_R^\alpha$ and $D_R^\alpha$ are the right-handed quarks with electric charges $\frac{2}{3}$ and $-\frac{1}{3}$, respectively, $e_R^\alpha$ are the right-handed charged leptons, $K$ is the Cabibbo-Kobayashi-Maskawa (CKM) matrix, $M_u$,  $M_d$ and $M_e$ are the diagonal mass matrices of the quarks and leptons, $M_W$ is the mass of the $W$-boson, and $\tilde{H}_i = \epsilon_{ij}H^{\dagger}_j$. \footnote{Summation over the repeated generation indices $\alpha$ and $\beta$ is understood in eq. (\ref{Yukawa}), with one exception: since the mass matrices $M_u$, $M_d$ and $M_e$ are diagonal, the notation $M^{\alpha\alpha}$ ($M^{\beta\beta}$) simply denotes their $\alpha$-th ($\beta$-th$\,$) diagonal entry, and no separate summation over the repeated Greek superscripts of the mass matrices is implied.}

The presence of the Higgs field shapes many aspects of the SM. The negative sign of the mass parameter in the Higgs potential leads to ``spontaneous breaking'' of the SU(2)$\times$U(1) symmetry down to electromagnetic U(1), the Higgs vacuum expectation value leads to masses of electroweak bosons $W$ and $Z$, to charged lepton masses, and ``current'' masses of quarks. 

 \subsection{The \nuMSM}
 \label{numsm1}
 A convincing experimental proof of the incompleteness of the Standard Model came at the end of the nineties and was associated with the observation of neutrino oscillations (for a comprehensive review and references to original experimental and theoretical works, see, e.g. \cite{Strumia:2006db}). It calls for the extension of the SM in the neutrino sector, and thus, to the existence of new leptonic states\footnote{For a review of alternatives to the type I seesaw mechanism see  \cite{Strumia:2006db}.}. 

The Standard Model has a suspicious asymmetry between the quark and lepton sectors. Namely, every left-handed quark or charged lepton has its right-handed counterpart. The neutral leptons -- neutrinos -- are treated differently in the SM; they do not have right-handed partners. This is exactly why neutrinos are massless in the SM and why there are three distinct conserved leptonic numbers. Therefore, it is natural to upgrade the SM by adding to it several right-handed neutral fermion states.  They can be called singlet or neutral leptons, sterile neutrinos, or alike. In the following, we will use the name ``Heavy Neutral Leptons'' (HNLs), following Particle Data Group. The theory with three HNLs -- the \nuMSM -- looks most natural as this is the number of fermionic generations in the SM. It also happens to be a minimal model, which enables us to explain simultaneously neutrino masses and oscillations, the presence of baryon asymmetry, and Dark Matter (DM) in the Universe \cite{Asaka:2005an,Asaka:2005pn}. 

The $\nu$MSM is described by the type I see-saw Lagrangian \cite{Minkowski:1977sc, Yanagida:1979as, Gell-Mann:1979vob, Mohapatra:1979ia, Schechter:1980gr}
\be
\mathcal{L_{\rm\nu MSM}} = \mathcal{L}_{\rm SM} + i\, \bar{N}_{I} \gamma^\mu \partial_\mu N_{I} - F_{\alpha I} \bar{L}_\alpha N_{I} \tilde{H} - \frac{M_{I J}}{2} \bar{N}_I^c N_J + h.c.\,,
\label{Lnumsm}
\ee
where $N_I$ ($I=1,~2,~3$) are the new neutrino states  and  $F_{\alpha I}$ are Yukawa couplings. Contrary to the quarks and charged leptons, getting their masses from Yukawa interaction with the Higgs field, the masses of HNLs (they are SU(3)$\times$SU(2)$\times$U(1) singlets) depend on an additional parameter -- the Majorana mass $M_{IJ}$. The value of Majorana mass (and, therefore, the value of the physical mass of an HNL) is not related to the Higgs vacuum expectation value and cannot be found in neutrino oscillation experiments.  The famous see-saw formula \cite{Minkowski:1977sc, Yanagida:1979as, Gell-Mann:1979vob, Mohapatra:1979ia, Schechter:1980gr} that relates the observed masses of the active neutrinos to the Majorana masses of HNLs does not allow us to fix even an order of magnitude of the HNL masses, which can be any number from a fraction of eV to $10^{15}$ GeV. The $\nu$MSM contains 18 new parameters in comparison with the SM. Three of them are the Majorana masses $M_{II}$ (the matrix $M_{IJ}$ can be chosen to be real and diagonal), while another 15 are hidden in the Yukawa matrix $F_{\alpha I}$  and can be chosen as 3 diagonal Yukawa couplings, 6 mixing angles, and 6 CP-violating phases. The active neutrino Majorana mass matrix is characterised by 9 combinations of these parameters (3 mixing angles and 1 Dirac phase in the PMNS mixing matrix, 3 neutrino masses, and 2 Majorana phases); many of them can be fixed by the experiments in neutrino physics (oscillations, neutrinoless double beta decay, and the measurements of the absolute scale of neutrino masses),  other \nuMSM parameters are only accessible with a direct discovery of HNLs.

Only these two theories - providing a self-consistent and minimal framework, and thus giving the maximal predictivity -  will be considered in what follows.

\section{Self-consistency of the SM and \nuMSM}
\label{consistency}
The use of the SM in cosmology requires its extrapolation to very high energies, possibly up to the Planck scale. So the question arises: can this be done in a consistent way? The SM (as well as the \nuMSM) is a renormalisable QFT, with only possible inconsistencies associated with the renormalisation group (RG) running of two coupling constants:  the Higgs self-coupling  $\lambda$, and the hypercharge  U(1) gauge coupling $g'$. First, both of them have the so-called Landau poles at a high energy scale, which may signal the inconsistency of the theory \cite{Landau:1954cxv} (see, however, \cite{Bogolyubov:1959vck}, where it was argued that this is simply an indication of a strongly coupled regime). The Landau pole in $g'$ occurs well above the Planck scale and is thus irrelevant - in any event, gravity cannot be neglected starting from energies $\sim M_P$. This is not so for $\lambda$. If the mass of the Higgs boson were larger than $170$ GeV, the Landau pole in the Higgs self-coupling would show up at energies below the Planck mass, indicating that, perhaps, the SM should be extended in some way at these scales. In any event, no perturbative computations would be valid above these energies.  Since the mass of the Higgs boson -- $m_H = 125.20 \pm 0.11$ GeV \cite{ParticleDataGroup:2024cfk} --  is well below this value, this scenario is not realised.

\begin{figure}[tb]
\centerline{\includegraphics[width=0.75\textwidth]{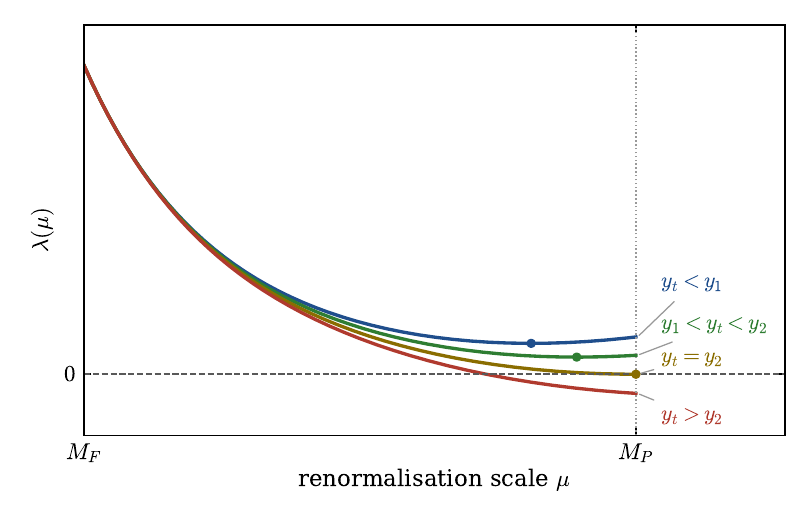}}
\caption{Schematic behaviour of the scalar self-coupling $\lambda$ as a function of the renormalisation scale $\mu$ for different values of the top quark Yukawa coupling $y_t$; $y_1$ and $y_2$ are defined in the text. All curves start from one and the same value of $\lambda$ at $\mu=M_F$, fixed by the measured Higgs boson mass. The dots mark the positions of the minima of $\lambda(\mu)$. For precise computations, see \cite{Bezrukov:2012sa, Degrassi:2012ry, Buttazzo:2013uya, Bednyakov:2015sca}.}
 \label{lambda}
\end{figure}

For the masses of the Higgs boson and of the top quark close to their experimental values, the behaviour of the scalar self-coupling is rather peculiar. The top quark loop contributes with a minus sign to the scalar self-coupling beta function, leading to a decreasing $\lambda$. At some point, other positive contributions to the running start to dominate, and $\lambda$ starts increasing.  As the mass of the Higgs boson is known with rather high accuracy, the main uncertainty in the $\lambda$ behaviour is associated with the top quark Yukawa coupling $y_t$. If it is smaller than a certain value, $y_1$, ``our'' minimum of the Higgs effective potential (which determines the Fermi scale,   $M_F=1/\sqrt{G_F} \approx 293$ GeV) is unique. If $y_1<y_t<y_2$, a second (metastable) minimum appears at a high energy scale. For $y_t<y_2$, our vacuum is absolutely stable. For $y_t>y_2$, the other minimum is deeper than ours, meaning that it is now ``our'' minimum which is metastable. Roughly speaking\footnote{The precise definition of $\lambda$ depends on the scheme of renormalisation.}, the latter situation corresponds to the case when $\lambda$ changes sign in some energy range, see Fig. \ref{lambda}.  

The structure of the effective potential happens to be very sensitive to the parameters of the SM. For instance, $(y_2-y_1)/y_1 \sim 10^{-5}$, corresponding to the change of the top quark mass by $\sim 0.5$ MeV. The vev of the Higgs field in the second minimum, if it exists, can vary between $10^{10}$ GeV and $\sim 10^{30}$ GeV if the mass of the Higgs boson and that of the top quark are varied within the admitted experimental values \cite{Andreassen:2014eha}.

If our vacuum were indeed metastable with a lifetime less than the age of the Universe, that would lead to the conclusion that the SM cannot be used for cosmology and that new physics must appear below the Planck scale. This situation, however, is not realised, as the experimental value of the Higgs mass is well above $\simeq 110$ GeV (for $m_t\simeq 172$ GeV) \cite{Altarelli:1994rb, Casas:1994qy, Casas:1996aq, Andreassen:2017rzq}, for which metastability would lead to a serious problem.

\begin{figure}[tb]
\centerline{\includegraphics[width=0.75\textwidth]{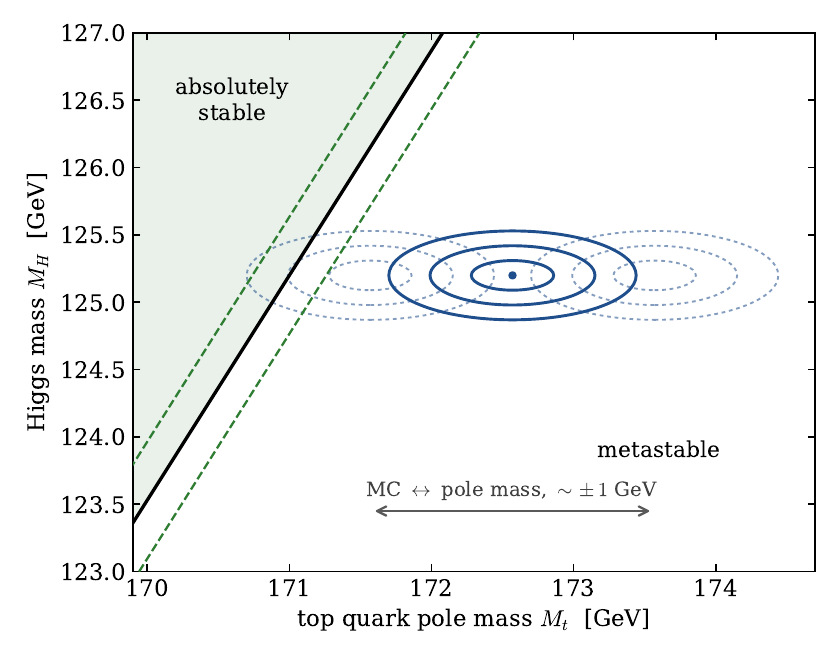}}
\caption{The borderline between the stable and metastable Higgs vacuum in the plane of the top quark pole mass and the Higgs mass. The solid line is the stability border, based on the most precise computation of Ref. \cite{Bednyakov:2015sca}; the green dashed lines correspond to the strong coupling constant $\alpha_s$ variation in the admitted experimental interval. Ellipses correspond to 1, 2, and 3 $\sigma$ standard deviations in the measurements of the Higgs mass and the top ``Monte-Carlo'' (MC) mass, centred at the PDG 2024 values $m_H = 125.20\pm 0.11$ GeV and $M_t = 172.57 \pm 0.29$ GeV \cite{ParticleDataGroup:2024cfk}. The dotted ellipses show the measured region displaced horizontally by $\pm 1$ GeV (horizontal arrow) -- the additional theoretical uncertainty in the relation between the MC and pole top masses, see the discussion of different definitions of t-quark mass in \cite{ParticleDataGroup:2024cfk}.} 
 \label{stability}
\end{figure}

Given the experimental uncertainties (dominated by the systematic errors in the relationship between the ``Monte-Carlo'' top quark mass and Yukawa coupling), we do not know whether our vacuum is absolutely stable or metastable, see Fig. \ref{stability} (for a review see \cite{Bezrukov:2014ina} and references therein). Let us explain why the situation remains undecided in spite of the precision of the measurements. The central ellipses in Fig. \ref{stability} do correspond to the currently measured values of the Higgs mass and of the top quark ``Monte-Carlo'' mass, and, taken at face value, they would favour metastability. However, the quantity which enters the stability analysis is the top quark Yukawa coupling in a well-defined renormalisation scheme (usually, $\overline{\rm MS}$), whereas the mass extracted from the kinematic fits at hadron colliders is a parameter of the Monte-Carlo event generators. The relation between the two carries an additional theoretical uncertainty of the order of $1$ GeV \cite{ParticleDataGroup:2024cfk}, larger than the experimental error bars. This uncertainty corresponds to a horizontal displacement of the ellipses in Fig. \ref{stability} by $\sim 1$ GeV in either direction, so that no definite conclusion about the (meta)stability of our vacuum can be drawn at present.

Despite these uncertainties, there are several statements which can be taken for granted:
\begin{itemize}
\item The parameters of the Standard Model are very close to the boundary separating the stable and metastable regions. This fact may find its explanation if the SM and gravity are asymptotically safe - a hypothesis put forward in \cite{Shaposhnikov:2009pv}, which led to the prediction of the Higgs boson mass before its discovery.
\item Even if our vacuum is metastable, its lifetime greatly exceeds the age of the Universe \cite{Andreassen:2014eha}\footnote{For the discussion of the EW vacuum decay in the presence of new physics and gravity see \cite{Branchina:2015nda, Branchina:2016bws, Bentivegna:2017qry, Branchina:2019tyy}. }. This conclusion is derived within the SM, i.e. under the assumption that no new physics modifies the running of $\lambda$ or the tunnelling rate between the Fermi and Planck scales. The question of metastability reopens in a non-trivial way when the SM is coupled non-minimally to gravity and the Higgs field is required to drive inflation; see the discussion of point 5 of the list in Section \ref{metricHI}.
\end{itemize}

Coming back to the question of whether it is self-consistent to use the SM  in the early Universe, the answer is ``yes'' if the EW vacuum is stable and ``maybe'' if it is metastable. In the latter case, an explanation should be given why the Universe ended up in the EW vacuum rather than in a more energetically favourable one with a large value of the Higgs vev, see Section \ref{metricHI}.

\section{Naturalness and hierarchy problems}
\label{Naturalness}
The presence of an elementary scalar field in the SM - the Higgs boson - raised yet another concern about the range of validity of the theory. Everybody knows that the loop Feynman diagrams are divergent when one integrates over the (loop) momentum, and thus, taken at face value, they do not make sense. These divergences can be consistently treated by renormalisation (see, e.g. textbooks \cite{Bogolyubov:1959bfo, Peskin:1995ev}). Still, the existence of power-like divergences in the Higgs sector creates a non-easy feeling.

The dominating philosophy of how to deal with this, possibly alarming, fact is based on two different hypotheses:

\begin{enumerate}[label=(\roman*)]
\item The first one states that at some energy scale, the ``cutoff'' $\Lambda$, the SM is replaced by some more fundamental theory.  There are little doubts that the cutoff does exist. The SM does not include gravity, and is therefore not a complete theory of forces in Nature. Gravity brings us a cutoff associated with the Planck mass $\Lambda \sim M_P \sim 10^{19}$ GeV. As the high energy experiments cannot touch energy domains above a few TeV, they do not exclude the existence of cutoffs below the Planck mass (but above several TeV). The prominent examples of theories with very high cutoffs include Grand Unified Theories (of strong, weak, and electromagnetic interactions) with $\Lambda \sim 10^{16}$ GeV \cite{Georgi:1974sy, Georgi:1974yf}, and theories with Peccei-Quinn symmetry and invisible axion, proposed to solve the strong CP-problem with $\Lambda \sim 10^9-10^{13}$ GeV \cite{Kim:1979if, Shifman:1979if, Zhitnitsky:1980tq, Dine:1981rt}. 
\item The second hypothesis is more technical and provides a recipe for estimating the divergent expressions in the SM. It states that the region of the loop integration over momenta should be constrained by the cutoff $\Lambda$. A support of this conjecture comes from the Wilsonian renormalisation group \cite{Wilson:1973jj}, providing a procedure of ``integrating out'' high energy physics above some (running) energy scale.
\end{enumerate}

If  (ii) is accepted (we will argue below that it should not be), one runs into troubles, which are dubbed ``fine tunings'' and ``naturalness''. Indeed, let us consider radiative corrections to the masses and couplings of different particles of the SM. Regularise the theory at high energies in a way which respects the gauge symmetries of the SM and the chiral structure of the fermions \footnote{This requirement is important. Gauge non-invariant regularisation would introduce the corrections $\propto \Lambda^2$ for gauge boson masses and $\propto \Lambda$ for fermions.}. Then, quantum corrections to all the parameters, except one - the mass of the Higgs boson, are only logarithmically divergent and are of the order $ \alpha/\pi^2 \log(\Lambda/M_F)$ (here $\alpha$  is the typical value of the square of the SM couplings), leading to the conclusion that the quantum effects are not so large for any cutoff scale below $M_P$.

This is not the case for the mass of the Higgs boson. The quantum corrections come from top quark, Higgs, and gauge loops. Several of these diagrams are {\em quadratically} divergent, leading to the most significant term $\delta m_H^2 \propto y_t^2\Lambda^2$. So, it looks like the ``natural value'' of the Higgs boson mass must be of the order of the cutoff scale. If the cutoff is as large as the Planck scale, to get the small observed Higgs mass, one has to ``fine-tune'' with astonishing accuracy the tree value of the mass $m_H^{\rm{tree}}$ to cancel the radiative corrections. So, one may conclude that the SM is highly fine-tuned and therefore ``unnatural''.  A similar statement is valid for the cosmological constant, which receives quartically divergent corrections, which should be tuned to the tree value even with much higher accuracy to get the experimentally observed value $\epsilon_{vac} = \left(2.24 \times 10^{-3} \mathrm{eV}\right)^{4}$.

The arguments about fine-tuning and naturalness look to be very convincing. They have motivated to search for ``natural'' extensions of the SM, where the cutoff scale (the onset of new physics) is right above the Fermi scale, so that the amplitude of the radiative correction to the Higgs mass is of the order of its observed value. The ``naturalness'' paradigm proposes to seek for BSM theories, in which the low-energy physics is organised in such a way that the small value of the electroweak scale should be stable against modifications of the ultraviolet (UV) physics and thus does not depend on the Grand Unified or Planck scales (for reviews see, e.g.  \cite{Giudice:2008bi, Giudice:2013yca, Craig:2023lyt}).  This new physics is supposed to ``screen'' the large UV scale from the low energy domain. Many BSM theories were invented with a value of the cutoff right above the Fermi scale. The examples include theories with dynamical electroweak symmetry breaking and the composite Higgs boson, supersymmetric (SUSY) theories with low energy SUSY breaking,  theories with large extra dimensions, all with $\Lambda \sim 10^3$ GeV (for reviews see, e.g. \cite{ParticleDataGroup:2024cfk}). In general, all these proposals lead to the prediction of new particles around the Fermi scale. Note also that if the same logic were applied to the cosmological constant puzzle, it would predict new physics at much smaller energies,  $\sim (\epsilon_{vac})^{1/4} \sim 10^{-3} \mathrm{eV}$. 

In addition to the ``naturalness'' criterion discussed above, a ``modified naturalness'' principle was put forward in \cite{Vissani:1997ys, Shaposhnikov:2007nj, Farina:2013mla}. It is inspired by dimensional regularisation, in which there are no power divergences. Still, if one uses the $\overline{\rm MS}$ subtraction scheme \cite{tHooft:1972tcz}  (throwing away poles in $\epsilon$, where $D=4-2\epsilon$ is the fractional dimensionality of the space-time), one finds that loops of superheavy particles (such as right-handed see-saw neutrinos with masses $\sim 10^{10-13}$ GeV), if present, ``give'' a contribution to the Higgs mass of the order of the superheavy particle mass.  This still requires fine-tuning between the parameters of the theory.

\rev{The physical significance of these finite ``threshold corrections'' is, however, questionable. As was shown in \cite{Mooij:2021ojy}, in renormalisable theories with well separated mass scales, all physics of the ``light'' sector can be computed in a way which does not involve any large contributions induced by the ``heavy'' sector, indicating that the apparent need for dramatic cancellations is an artefact of the conventional computational procedure rather than a property of the theory itself; see the discussion of the finite QFT formalism below.}

If the ``naturalness'' ideas were right, the study of the cosmological consequences of the Standard Model (and of the Higgs boson) at temperatures exceeding a few TeV would not make much sense. Indeed, in this case, new physics would dominate, or at least significantly influence, the evolution of the Universe. 

Let us now discuss whether the ``naturalness'' criterion, based on hypothesis (ii), should be accepted. From the experimental side, this concept has not gained any support.  The experiments at the high-energy frontier carried out at CERN provided a beautiful confirmation of the Standard Model without any convincing signs of new physics. The searches for new physics have not produced any positive results, excluding the domains of theoretical prejudice based on different  ``natural'' models of new physics. Moreover, naturalness does not work with the cosmological constant - we do not see new physics around a fraction of eV.

Theoretically,  the ``naturalness'' paradigm is questioned by the following simple observation. Any meaningful theory aims to provide a mapping of a few parameters -- fundamental constants -- into a multitude of physical predictions. This is the case, for example, in Quantum Electrodynamics. By knowing the fine structure constant $\alpha$ and the electron mass $m_e$ (which can be determined in two experiments), we can predict the outcome of an infinite number of different experiments. Clearly, neither the basic parameters $\alpha$ and $m_e$ nor the results of any experiment are divergent quantities - they are all some finite numbers.  This means that all the divergences in Feynman diagrams are artificial and related to a particular formalism - multiplicative renormalisation, used in practically all QFT computations. In this formalism, the divergences are dealt with by first regularising the Feynman diagrams (say, by introducing a UV cutoff, or using dimensional regularisation), and then by removing the divergences by adding to the theory specially constructed counter-terms, ``fine-tuning'' them to avoid infinities in the final result. These ``bare'' quantities --  masses and coupling constants -- are infinite when taking away the regularisation, and extremely ``fine-tuned''. This fine-tuning, however, is unphysical, as it happens between quantities which cannot be measured in experiment.  

Therefore,  there must exist a procedure of mapping between physical (finite) parameters  {\em without encounter of any divergent expressions at any stage of the computation}.  If this all-way-finite QFT formalism is used, there are no divergent Feynman diagrams whatsoever, all expressions are finite, and the computation of low-energy observables does not require knowledge of the UV domain of the theory. No ``fine-tuning'' is needed between the ``tree value'' of the Higgs mass and radiative corrections. The argument in favour of supersymmetry or composite Higgs models based on the absence of quadratic divergences becomes diluted, since no divergences exist at all in the finite QFT formulation.

A detailed discussion of the construction of a finite QFT  formalism can be found in \cite{Mooij:2021ojy, Mooij:2021lbc, Mooij:2024rys} and references therein. It can be applied to theories with several distinct mass scales, and even to non-renormalisable theories \cite{Ageeva:2024qie}.  The use of finite QFT methods demonstrates that theories like the stand-alone Standard Model,  the \nuMSM even with superheavy Majorana fermions, GUTs with or without supersymmetry, and these theories with gravity added as an extra interaction, do not have physically significant fine-tunings. The mass of the Higgs boson in these theories is an input parameter; its ``tree value''  cannot be separated from a radiative correction to it, so it does not make sense to discuss a fine-tuning between them.

To the best of the author's knowledge, the finite-QFT resolution of the fine-tuning puzzle described above has not been challenged in any publication. For a map of the different positions, the reader may consult the reviews \cite{Giudice:2008bi, Giudice:2013yca, Craig:2023lyt}.

A final comment concerns the theories where the mass of the Higgs boson can be ``computed in terms of more fundamental parameters''.  This definition is not very precise, as no unique answer to the question ``Which parameter is more fundamental than the others?'' exists. Moreover, the concept of computability is more philosophical than physical -- though we wish to have a completely predictive theory without free parameters, it is far from being guaranteed that this is the case.  Also, why only the Higgs boson mass?  We can equally require ``computability'' of the cosmological constant, the fine structure constant, the top quark mass, the $\Lambda_{QCD}$ scale of quantum chromodynamics, etc. Incidentally, quantum gravity effects, associated with tunnelling transitions with the change of spatial topology, make all of the parameters of the SM uncomputable even in principle, see  \cite{Coleman:1988cy,Coleman:1988tj} and a review \cite{Hebecker:2018ofv}.

Currently,  there are three classes of theories where the Higgs mass is said to be ``predictable''. These are theories with a composite Higgs boson, extensions of the SM based on dynamically broken SUSY, and large extra dimensions. The original motivation for these theories came from the desire to cut the quadratic divergences in the Higgs mass at the Fermi scale. As we have discussed, this concern is not valid in the finite QFT formulation, being thus formalism-dependent and therefore unphysical. In these three types of models, generically,  the existence of a multitude of new particles with masses of hundred GeV-TeV is predicted.  The absence of new physics in experiments at LHC and elsewhere requires fine-tuning of the parameters of these models, making them “unnatural”, whereas the attempts to save them require substantial machinery.

To summarise,  the author of this review does not see a theoretical motivation for the existence of new physics right above the Fermi scale inspired by ``naturalness''. So, it does make sense to study the cosmological consequences of the SM or the \nuMSM, using the fact that they are self-consistent QFTs, valid up to very high (possibly Planck) scales. In any event, this is the most predictive approach. It is described below. 

\section{From the symmetric phase to the Higgs phase at the electroweak scale} 
\label{EWPT}
 The ground state of the SM is determined by the minimisation of the effective potential of the Higgs field, which in tree approximation is given by eq. (\ref{pot}), leading to
 \be
 \label{zeroTvev}
 v^2  = \frac{ m^2}{\lambda}\approx (250~ {\rm GeV})^2\,,
\ee
where $v$ is the vacuum expectation value of the Higgs field, $H=(0,v/\sqrt{2})$. It is said that the EW symmetry is broken or that the system finds itself in the Higgs phase. The mass of the Higgs boson is $m_H^2=2\lambda v^2=2m^2$.
 
In the early Universe, the ground state of the SM is determined by the system's free energy (with a minus sign) considered as a function of the classical scalar field and the temperature $T$.  Accounting for one-loop quantum corrections to the effective potential at non-zero temperatures leads to the modification of (\ref{pot}) \cite{Kirzhnits:1972ut, Weinberg:1974hy}. If the temperature $T$ is much larger than the masses of all the particles in the SM, the leading temperature corrections to the effective potential scale as $T^2$ and modify the mass parameter $m^2$.  The effective potential in this domain of temperatures is given by
\be
V= m^2(T)  H^\dagger H + \lambda (H^\dagger H)^2\,,
\label{Tnot0}
\ee
where
\be
m^2(T) =  -m^2 +\kappa T^2\,,
\label{m3}
\ee
and
\be
\kappa=\frac{1}{2}\lambda+\frac{3}{16}g^2+\frac{1}{16}g'^2+
\frac{1}{4}y_t^2\,.
\label{kappa}
\ee
The four terms in eq. (\ref{kappa}) originate from the one-loop thermal corrections induced by the Higgs self-interaction, the SU(2) and U(1) gauge boson loops, and the top quark loop, respectively \cite{Kirzhnits:1972ut, Weinberg:1974hy}. The top quark dominates among the fermions because of its large Yukawa coupling; the contributions coming from the loops of charged leptons and lighter quarks are proportional to the squares of the corresponding Yukawa couplings, and thus are numerically small and can be omitted.

Since $\kappa>0$, at sufficiently large temperatures, the mass parameter of the effective potential changes sign at 
\be
T_c^2=\frac{m_H^2}{2\kappa}\,.
\label{H_c}
\ee
It is often said that at this temperature the EW symmetry is restored or that the system is in the symmetric phase. Indeed, the minimisation of the potential (\ref{Tnot0}) would give $v=0$ at $T>T_c$. The main reason for this ``symmetry restoration'' is the thermal fluctuations of the Higgs field, leading to $\langle H^\dagger H\rangle_T \sim  T^2$.

When the Universe cools down from high to lower temperatures, phase transitions might occur in the vicinity of $ T \simeq T_c$. First-order phase transitions and rapid changes in the systems' parameters near second-order phase transitions or cross-overs may lead to observable consequences. Assume for a moment that we have a first-order phase transition. It goes through the bubble nucleation in the same way as the water boils or the vapour condenses, and thus induces significant deviations from thermal equilibrium.  During the process of cooling of the Universe, the system finds itself in a state where $\phi$ is still zero even for $T$ somewhat smaller than $T_c$ because the probability of bubble nucleation with a new phase inside is suppressed at $T \approx T_c$. Thus, the Universe is in a super-cooled state, the decay of which may be quite a violent event consisting of the nucleation of bubbles of a new phase, their expansion, and the collision of different bubbles, which finally leads to the disappearance of the old phase. 

The first-order phase transition is an important ingredient of electroweak baryogenesis \cite{Shaposhnikov:1987tw, Shaposhnikov:1987pf, McLerran:1990zh, Cohen:1990it, Cohen:1990py, Turok:1990in, Dine:1990fj, Nelson:1991ab} (for the most recent review see \cite{vandeVis:2025efm}). Also, the electroweak first-order phase transition was discussed in connection with the generation of cosmological magnetic fields, primordial black hole formation, and the production of primordial gravitational waves; see \cite{Vachaspati:2020blt, Caprini:2009yp, Caprini:2019egz} and \cite{Caprini:2025trt} for a review. 

It is important to understand that the existence of phase transitions is not guaranteed, as one might think from using the term ``broken symmetry''. Gauge symmetries always stay intact \cite{Elitzur:1975im}, and there is no local gauge-invariant order parameter which can distinguish the symmetric and Higgs phases \cite{Banks:1979fi, Fradkin:1978dv}. In other words, the widely used terms ``symmetry breaking'' and ``symmetry restoration'' -- which we adopt throughout this review as a convenient and commonly accepted shorthand -- do not have an invariant meaning: only gauge-invariant quantities, such as the Higgs condensate $\langle H^\dagger H\rangle_T$ appearing below in the discussion of the electroweak crossover, characterise the state of the system. Thus, from these general considerations, the Universe may go through a first or second order phase transition, or simply gradually change its bulk characteristics, going through a smooth ``crossover''.

\begin{figure}[tbh]
\centerline{\includegraphics[width=8.5cm]{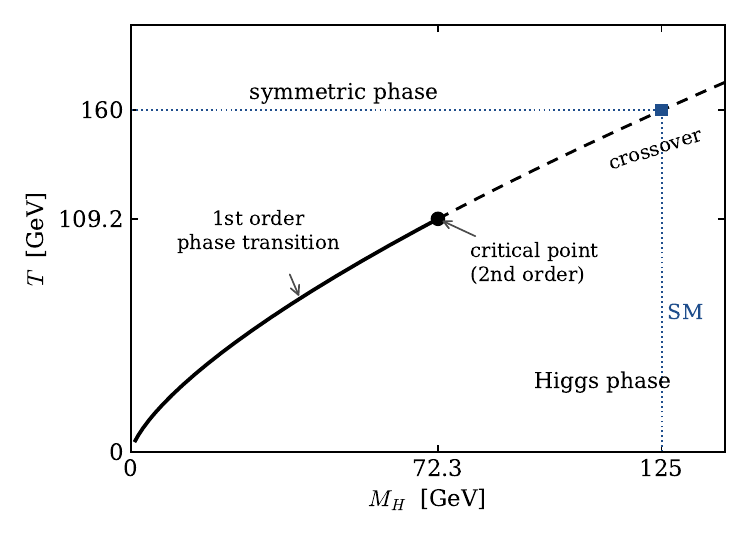} }
\caption{The schematic phase diagram of the Standard Model in the plane of the Higgs mass and the temperature. The position of the critical point and the crossover temperature for the physical value of the Higgs mass are taken from the lattice studies \cite{Kajantie:1996mn, DOnofrio:2015gop, Rummukainen:1998as}.} 
\label{ewphase} 
\end{figure}

What happens depends on the parameters of the theory. Fixing all of them besides the mass of the Higgs boson, perturbative and non-perturbative studies \cite{Farakos:1994kx, Kajantie:1996mn, Kajantie:1995dw, Kajantie:1995kf, Fodor:1994dm,  Fodor:1994sj, Laine:2012jy,  DOnofrio:2015gop} reveal the phase diagram shown in Fig. \ref{ewphase}.

The position of the critical point (the end-point of the first-order phase transition line)  is given by $T= 109.2 \pm 0.8~\mbox{GeV},~ m_H = 72.3 \pm 0.7~\mbox{GeV}$. Near this point, the SM is in the same universality class as the 3D Ising model \cite{Rummukainen:1998as}.

The analysis, which has led to elucidating the phase structure of the SM, is rather involved. At first sight, it looks like everything can be done analytically, since the couplings of the SM are all small at the electroweak scale.  And, indeed, this is the case for the left side of Fig. \ref{ewphase}, when the mass of the Higgs boson is sufficiently small. In this case, the phase transition is of first order, the jump of the order parameter (vev of the Higgs field) is sufficiently large, and perturbation theory can be trusted. This is not the case for that part of the phase diagram in the vicinity of the critical point. In fact, the expansion parameters at high temperature also contain the occupation numbers of bosons at small momentum transfer and read $\alpha T/m$ \cite{Linde:1980ts}, where $m$ is a typical mass of bosonic excitations in plasma. Thus, in the vicinity of the critical point, where the mass of the scalar excitation is small, the perturbation theory diverges badly. The physics of strongly interacting modes can be isolated. It can be shown that it is described by a {\em three-dimensional} purely bosonic theory, containing just the SU(2)$\times$U(1) gauge bosons and the Higgs doublet, with parameters perturbatively computable in terms of physical observables (such as masses of different particles and other parameters of the SM, and temperature) \cite{Farakos:1994kx, Kajantie:1995dw}.

Now, this effective three-dimensional theory can be studied non-perturbatively with the use of high precision lattice simulations \cite{Kajantie:1995kf, DOnofrio:2015gop}. They allow to extract many relevant physical observables associated with the transition from the symmetric to the Higgs phase (such as the latent heat of the first-order phase transition if it takes place, evolution of different correlation lengths, etc.)

As the mass of the experimentally observed Higgs boson ($125$ GeV) is greater than that corresponding to the critical point, there is no phase transition in the SM.  The electroweak crossover in the SM happens at temperature $T\simeq 160$ GeV (there is no exact definition of the crossover temperature since the thermodynamical characteristics of the system are continuous functions of $T$; the quoted value corresponds to the temperature at which the Higgs scalar condensate $\langle H^\dagger H\rangle$ changes most rapidly).

The phase structure of the \nuMSM and the position of the critical point are the same as those of the SM. If HNL's have masses well above the Fermi scale, their density is Boltzmann suppressed at temperatures $\sim 160$ GeV.  If their masses are of the order of the Fermi scale or below, their Yukawa couplings to the SM fermions and the Higgs boson are required to be very small (this comes from the see-saw formula). Thus, in both cases, they have no influence on the dynamics of gauge and scalar fields at temperatures of the order of the Fermi scale.  

The electroweak baryogenesis does not happen, nor can gravitational waves nor magnetic fields be generated in the SM and \nuMSM at the crossover temperature.

Extensions of the SM in the Higgs sector or containing higher-dimensional operators may undergo a first-order phase transition in some portion of their parameter space (see, e.g. \cite{Laine:2012jy, Bruggisser:2018mrt}). There is a rich literature discussing baryogenesis, gravitational waves and magnetic fields in these theories, see, e.g. \cite{Cohen:1993nk, Rubakov:1996vz, Trodden:1998ym, Riotto:1999yt, Dine:2003ax,  Morrissey:2012db, Garbrecht:2018mrp,  Bruggisser:2022rdm, Domcke:2025itc, Bhusal:2025lvm, vonHarling:2023dfl}.

\section{CP-violation in Higgs-fermion interactions and baryon asymmetry of the Universe}
\label{BAU}
\subsection{The SM}
\label{bausm}
Up to now, we have been discussing mainly the Higgs potential. Now we turn to yet another important part of the Higgs interactions, namely those described by the Yukawa terms in Eqns. (\ref{Yukawa},\ref{Lnumsm}). 

Historically, before the discovery of atmospheric neutrino oscillations in 1998 \cite{Super-Kamiokande:1998kpq}, there were no signs of new physics beyond the SM, and it was appropriate to ask at that time whether the SM alone could produce baryon asymmetry of the Universe (BAU) \cite{Shaposhnikov:1987tw}. Potentially, it has all the ingredients to do so. There is rapid baryon and lepton number violation at high temperatures \cite{Kuzmin:1985mm} due to the electroweak anomaly \cite{Adler:1969gk, Bell:1969ts, tHooft:1976rip}. The deviations from thermal equilibrium may be provided by the expansion of the Universe, whereas the presence of CP-violation is an experimental fact (these are the three Sakharov conditions for baryogenesis \cite{Sakharov:1967dj}).

The baryon and lepton number violation in the SM is associated with non-perturbative transitions between different topological vacua in the SU(2) gauge sector of the SM (for a review, see \cite{Rubakov:1996vz}).  At zero temperatures, these transitions are described by quantum-mechanical tunnelling and are exponentially suppressed \cite{tHooft:1976snw}, making them unobservable in particle physics experiments. At non-zero temperatures, they are associated with thermal fluctuations overpassing the energy barrier separating different topological vacua \cite{Kuzmin:1985mm}. They are called ``sphaleron'' transitions, the name coming from a specific configuration of the gauge and Higgs fields - an unstable solution to the classical equations of motion \cite{Klinkhamer:1984di}, corresponding to the top of the barrier. Non-perturbative lattice simulations allowed to determine precisely the temperature at which the sphaleron transitions in the SM are in thermal equilibrium in the expanding Universe \cite{DOnofrio:2014rug}, $T>T_{\rm sph}$, $T_{\rm sph}\approx 130$ GeV. 

It was realised already in \cite{Shaposhnikov:1986jp} that the structure of Higgs-fermion interactions in the SM makes the effects of CP violation in the processes with baryon number non-conservation very suppressed.  In the SM, the Yukawa terms (\ref{Yukawa}) provide a unique source of CP violation\footnote{We neglect the CP-violation which may come from non-trivial topological structure of the QCD vacuum, associated with the so-called $\theta$ term. Even if present, its effects at temperatures $T \gtrsim 160$ GeV, where baryon number is not conserved in the SM, are too small \cite{Kuzmin:1992up}.}.  The structure of CP violation in the SM is very peculiar. If a pair of quarks with charges $2/3$, or a pair of quarks with charges $-1/3$, or one of the mixing angles between different generations were zero, the SM would acquire an extra global symmetry which would allow to rotate away the complex Kobayashi-Maskawa phase. In other words, CP violation in the SM vanishes together with the combination
 \begin{eqnarray}
\label{jarlskog_det}
 D_{CP} &=& \sin(\theta_{12}) \sin (\theta_{23}) \sin(\theta_{13}) \sin
\delta_{CP} \nonumber \\
 && \cdot\, (m_t^2 - m_c^2)(m_t^2 - m_u^2)(m_c^2 - m_u^2)(m_b^2 -
m_s^2)(m_b^2 - m_d^2)(m_s^2 - m_d^2)~, 
\end{eqnarray}
known as the Jarlskog determinant \cite{Jarlskog:1985ht} (here $\delta_{CP}$ is the KM phase, and the Particle Data Group parametrisation of the CKM matrix is used.) A relevant dimensionless measure of CP violation in the Universe at the temperature where the sphaleron transitions are still active would be \cite{Shaposhnikov:1987tw} $\delta^{CP}_{KM} \sim D_{CP}/T_{\rm sph}^{12} \sim 10^{-20}$, where the quantity $D_{CP}$ of eq. (\ref{jarlskog_det}), of mass dimension twelve, is normalised by the corresponding power of the temperature, evaluated at the sphaleron freeze-out temperature $T_{\rm sph}\approx 130$ GeV introduced above. The result is some ten orders of magnitude smaller than the baryon asymmetry of the Universe, $n_B/s \simeq 8.7\times 10^{-11}$, where $n_B$ and $s$ are the baryon number and entropy densities in the Universe, respectively.  So, unless an enhancement of CP  violation is found, the baryon asymmetry of the Universe cannot be explained in the SM. Several attempts to find a mechanism overcoming this suppression were made in \cite{Shaposhnikov:1987tw, Shaposhnikov:1987pf, Nasser:1994bj, Farrar:1993hn, Kharzeev:2019rsy}, but all of them failed because of this or that reason \cite{Ambjorn:1988gf, Gavela:1994dt, Huet:1994jb, Hong:2023zrf}. 

Yet another theoretical result adds extra confidence to the statement that the SM cannot be responsible for the baryon asymmetry of the Universe.  BAU may be generated only at the temperatures $T>T_{\rm sph}$, when the processes with baryon number non-conservation are active.  At this time, all {\em perturbative} particle reactions are well in thermal equilibrium. They keep the distributions of quarks and leptons in equilibrium, and, because of this, the production of any nonnegligible baryon asymmetry is greatly suppressed \cite{Dolgov:1979mz}. A possible way out is to have a first-order electroweak phase transition. However, it does not happen as was discussed in Section \ref{EWPT}, meaning that we do need new physics to explain the baryon asymmetry of the Universe.

\subsection{The \nuMSM }
\label{baunumsm}
Going from the SM to the \nuMSM completely changes the baryogenesis story. First, in the SM, there are three exactly conserved anomaly-free fermionic numbers, $L_\alpha-\frac{1}{3} B$, where  $B$ is the baryonic number. The \nuMSM does not have any global symmetries at all, and thus no conserved fermionic numbers -- the lepton symmetries are broken by the Majorana mass terms for HNLs. Second,  the type I seesaw Lagrangian (\ref{Lnumsm}) has a very rich CP-violating structure: it contains 6 complex phases in the leptonic sector, instead of just one (the Kobayashi-Maskawa phase) in the SM.  Third, the presence of new particles -- the HNLs -- provides a new source of thermal nonequilibrium  - the HNLs can ``freeze in'' and ``freeze out'' during the evolution of the Universe. 

Contrary to the SM, the \nuMSM contains a dark matter candidate - the lightest HNL $N_1$, often called in this context ``sterile neutrino DM''  (for a short overview see Section \ref{DM}).  This DM candidate was proposed in \cite{Dodelson:1993je} and is the subject of numerous studies (for reviews see, e.g. \cite{Boyarsky:2009ix, Kusenko:2009up,  Boyarsky:2018tvu}).  If, indeed, all the DM in the Universe is in the form of sterile neutrinos, several important consequences for particle physics can be drawn. 

The Yukawa interactions of $N_1$ to the leptons of the SM must be very small, to ensure the stability of $N_1$ at cosmological scales. This means that the HNL does not contribute to the seesaw formula \cite{Asaka:2005an, Boyarsky:2006jm},  and one of the active neutrino states is effectively massless, thereby fixing the overall active neutrino mass scale \cite{Asaka:2005an, Boyarsky:2006jm}. In addition, the number of parameters of the \nuMSM relevant for neutrino masses and baryogenesis is substantially reduced. Instead of 18 parameters for three HNLs, the theory where the DM sterile neutrino is dropped from the Lagrangian contains 11 new parameters in comparison with the SM: 2 Majorana neutrino masses of HNLs,  2 Dirac neutrino masses, 4 mixing angles and 3 CP-violating phases. Seven combinations of these parameters are hidden in the active neutrino Majorana mass matrix (since the smallest neutrino mass is negligibly small, putting it to zero eliminates one Majorana phase from it, reducing the number of low-energy parameters from 9 to 7).

Qualitatively, baryogenesis may happen at freeze-out or freeze-in of HNLs. In the first case, HNLs spend some time in thermal equilibrium and then go out of it (freeze-out)  at temperatures of the order of, or above, their mass.  Later, they decay into a lepton and the Higgs boson via Yukawa couplings (\ref{Lnumsm}) \cite{Fukugita:1986hr}. These processes produce lepton asymmetry of the Universe, converted into baryon asymmetry in sphaleron reactions. In the second case, HNLs enter into thermal equilibrium at temperatures of the order of the Fermi scale and produce lepton asymmetry in HNL oscillations \cite{Asaka:2005pn, Akhmedov:1998qx}. In these processes, both the Higgs field background and the Higgs particle play an important role.  Again, this asymmetry is transferred into baryon asymmetry in sphaleron transitions. 

In fact, both mechanisms (freeze-in and freeze-out) are described by the same kinetic equations and allow for systematic study without any simplifying assumptions \cite{Klaric:2020phc}. The bottom line of studies of baryogenesis in the theory with 2 essential HNLs can be summarised as follows. All the data on neutrino oscillations can be reconciled with successful baryogenesis, without running into any drawback with Big Bang Nucleosynthesis, for the HNL masses as small as $140$ MeV and as large as $\sim 10^{10}$ GeV. If all three HNLs contribute to neutrino masses and participate in baryogenesis, the admissible parameter space is extended. In particular, the mixing angles between HNLs and active neutrinos can be substantially larger than in the two HNL case \cite{Abada:2018oly, Drewes:2024bla}.

The indicative ranges of the \nuMSM parameters -- the HNL masses, Yukawa couplings, and active-sterile mixing angles -- that can account for the baryon asymmetry and for Dark Matter, while being consistent with the existing experimental and cosmological constraints, are collected in Table \ref{tab:numsm}. A comment on the magnitude of the Yukawa couplings is in order. The see-saw relation ties the active neutrino masses to the combination $F^2 v^2/M$, so that for HNLs with masses at or below the Fermi scale the Yukawa couplings must be tiny, $|F| \sim \sqrt{m_\nu M}/v \sim 10^{-8}\, (M/1\,{\rm GeV})^{1/2}$, i.e. below  the electron Yukawa coupling $y_e \simeq 3\times 10^{-6}$ (unless there is a symmetry ensuring cancellations in the see-saw formula, \cite{Shaposhnikov:2006nn}).  This is at variance with the original spirit of the see-saw mechanism \cite{Minkowski:1977sc, Yanagida:1979as, Gell-Mann:1979vob, Glashow:1979nm, Mohapatra:1979ia}, where the Yukawa couplings were {\em assumed} to be of the order of those of the quarks and charged leptons, and the smallness of the neutrino masses was attributed to Majorana masses close to the GUT scale.

\begin{table}[tb]
\caption{\label{tab:numsm}Indicative ranges of the \nuMSM parameters for which the theory can explain the observed baryon asymmetry of the Universe and Dark Matter, in consistency with neutrino oscillation data, Big Bang Nucleosynthesis, direct searches, X-ray observations, and structure formation. The quoted numbers are approximate and correspond to the benchmark scenario in which $N_1$ is the Dark Matter particle, while the nearly degenerate pair $N_{2,3}$ generates the active neutrino masses and the baryon asymmetry; see \cite{Boyarsky:2009ix, Boyarsky:2018tvu, Klaric:2020phc, Abada:2018oly, Drewes:2024bla} and the text for details, and Fig.~14 of \cite{Boyarsky:2018tvu} together with the sensitivity studies \cite{Alekhin:2015byh, FCC:2025lpp} for the detailed exclusion plots. For comparison, the electron Yukawa coupling is $y_e\simeq 3\times10^{-6}$.}
\footnotesize
\begin{tabular}{@{}llll}
\hline\hline
& $N_1$ (Dark Matter) & \multicolumn{2}{l}{$N_{2,3}$ (baryogenesis and neutrino masses)}\\
\hline
Majorana mass & $M_1 \sim 1-50$ keV & \multicolumn{2}{l}{$140~{\rm MeV} \lesssim M_{2,3} \lesssim 10^{10}$ GeV, $M_2\simeq M_3$}\\[2pt]
Yukawa couplings $|F_{\alpha I}|$ & $\lesssim 10^{-11}$ & \multicolumn{2}{l}{$\sim 10^{-8}-10^{-4}$, scaling as $\sqrt{m_\nu M_{2,3}}/v$}\\[2pt]
Mixing angle $U^2$ & $\sin^2(2\theta_1) \sim 10^{-13}-10^{-7}$ & \multicolumn{2}{l}{from the see-saw line $U^2\sim 5\times 10^{-11}\left(\frac{\rm GeV}{M_{2,3}}\right)$}\\
& (anti-correlated with $M_1$) & \multicolumn{2}{l}{up to $\sim 10^{-6}-10^{-5}$ for $M_{2,3}$ in the GeV range}\\[2pt]
Main constraints & lifetime, X-rays, DM abundance, & \multicolumn{2}{l}{neutrino oscillations, BBN, direct searches,}\\
& structure formation & \multicolumn{2}{l}{requirement of successful baryogenesis}\\
\hline\hline
\end{tabular}
\end{table}

If the HNL masses are smaller than the mass of the $Z$-boson, the see-saw mechanism of neutrino mass generation and baryogenesis can be verified experimentally \cite{Shaposhnikov:2006nn, Gorbunov:2007ak, Blondel:2014bra, Hernandez:2016kel, Drewes:2021nqr, Hernandez:2022ivz, Sandner:2023tcg} at new experiments at CERN, such as SHiP \cite{Alekhin:2015byh}, and later at FCC-ee \cite{FCC:2025lpp} or LEP-3 \cite{Anastopoulos:2025jyh}, or LEP-Z \cite{Drewes:2025ohy}.

\section{The Higgs boson and Dark matter}
\label{DM}
The SM does not contain a candidate for a DM particle, whereas \nuMSM has - this is the lightest HNL, $N_1$. Of course, a vast landscape of alternative particle DM candidates exists -- from WIMPs to axions to primordial black holes -- which lie outside the minimal framework adopted in this review; for comprehensive overviews, see \cite{Bertone:2004pz, Cirelli:2024ssz}. If this particle has no other interactions beyond (\ref{Lnumsm}), the Higgs boson Yukawa couplings $F_{\alpha I}$ with $N_1$ determine the relic abundance of the DM as well as the decay rate of $N_1$.  The parameters $F_{\alpha I}$ cannot be found within the \nuMSM, as is the case with the parameters of the SM.  The following considerations constrain a possible range of masses and couplings of the DM sterile neutrino ~\cite{Dodelson:1993je, Boyarsky:2009ix, Boyarsky:2018tvu, Shi:1998km, Abazajian:2001nj, Abazajian:2001vt, Abdullahi:2022jlv}:
\begin{enumerate}
\item The life-time of $N_1$ must exceed the age of the Universe. Moreover, $N_1$ can decay into a photon and a neutrino, producing a narrow X-ray line which can be detected by the X-ray telescopes in space (such as XMM-Newton,  Chandra, or XRISM). This results in an upper limit on the sterile neutrino mixing angle with active neutrinos (see \cite{Boyarsky:2018tvu} and references therein), which depends on the $N_1$ mass  $M_1$.
\item Yet another upper bound on this mixing angle comes from the cosmological production of sterile neutrinos.  The typical temperature of the cosmological production of sterile neutrinos is $\sim 200$ MeV.  If $N_1$ interact too strongly,  the reactions like $l^+l^-\to N_1\nu$ would over-produce them, making the abundance of Dark Matter in the Universe larger than observed. This limit is also mass-dependent.
\item A lower bound on the mixing angle also comes from cosmology. If it is too small, the \nuMSM interactions cannot produce enough  DM. This limit is considerably less certain than the upper limits (1) and (2), and depends on the lepton asymmetry of the Universe \cite{Shi:1998km} (for the most recent studies see \cite{Gorbunov:2025nqs, Akita:2025txo, Vogel:2025aut}) and details of a possible QCD phase transition \cite{Shaposhnikov:2023hrx}. The requisite lepton asymmetry may be produced in the \nuMSM interactions below the sphaleron freezout \cite{Shaposhnikov:2008pf, Canetti:2012vf, Canetti:2012kh, Ghiglieri:2020ulj}. In addition, the lower bound is also sensitive to the way gravity couples to the \nuMSM. For instance, in the Einstein-Cartan formulation of gravity (discussed in detail in Sections \ref{EC} and \ref{Weyl} in the context of Higgs inflation), a new four-fermion interaction of $N_1$ results in its efficient high-temperature production \cite{Shaposhnikov:2020aen}. In the latter case, the lower bound completely disappears, whereas the structure formation bounds can be easily satisfied. This gravitational production mechanism ties the Dark Matter discussion of the present Section to the choice of the gravitational formulation, which is the subject of Section \ref{HI}.
\item Several bounds on the mass of $N_1$ come from structure formation. The Tremaine-Gunn bound \cite{Tremaine:1979we} on the mass of any fermionic DM particle, which comes from the Pauli exclusion principle applied to the dwarf galaxies, puts a robust lower bound on $M_1$, $M_1\gtrsim 1$ keV (see \cite{Bezrukov:2025ttd} for a most recent study). The analysis of the Lyman-$\alpha$ forest data allows us to constrain the free streaming length of the light DM particles, which can be converted into the mass bound, provided the mechanism of the sterile neutrino production is fixed \cite{Boyarsky:2018tvu, Viel:2005qj, Boyarsky:2008xj, Viel:2013fqw}.  In addition, the number of Milky Way satellite galaxies \cite{DES:2020fxi, Enzi:2020ieg, Newton:2024jsy}) also depends on the mass and velocity distribution of the DM particles, providing extra limitations.
\end{enumerate}
The summary of different constraints can be found in Fig. 14 of Ref. \cite{Boyarsky:2018tvu}; for the description of the uncertainties involved, see ~\cite{Garzilli:2019qki, Garzilli:2018jqh}. The bottom line is that a sterile neutrino with the mass in the ${\cal O} (10)$ keV range and sufficiently small mixing angle can be a viable DM candidate, satisfying all current constraints.

\section{The Higgs boson as the inflaton}
\label{HI}
Cosmological inflation is the leading paradigm for describing the earliest observable stage of the evolution of the Universe.  It is based on the hypothesis of having a period of accelerated expansion preceding the radiation-dominated epoch. This idea explains, in a unified way, the observed large-scale homogeneity and isotropy of the Universe, the near-flatness of spatial geometry  \cite{Starobinsky:1980te, Guth:1980zm, Linde:1981mu, Albrecht:1982wi}, and the origin of primordial perturbations that seeded cosmic structure formation \cite{Mukhanov:1981xt}.

The simplest realisation of inflation invokes a scalar field slowly evolving in a sufficiently flat potential \cite{Linde:1983gd}. The scalar field provides an approximately constant vacuum energy density, driving accelerated expansion, while its quantum fluctuations seed primordial inhomogeneities. This raises central questions for particle physics: ``What is the identity of the inflaton? Is this a new particle yet to be discovered?''

From the perspective of particle physics, inflation requires a scalar degree of freedom with two essential properties. First, its potential must be sufficiently flat to sustain slow-roll evolution. Second, the inflaton must couple to Standard Model degrees of freedom strongly enough to transfer its energy into ordinary particles after inflation, thereby reheating the Universe and initiating the Hot Big Bang.

In many inflationary models, these requirements motivate the introduction of new scalar fields beyond the SM. The representative examples include natural inflation, driven by a pseudo-Nambu-Goldstone boson \cite{Freese:1990rb}, and its numerous axion-inflation descendants (for a review, see \cite{Pajer:2013fsa}), the large class of $\alpha$-attractor models \cite{Kallosh:2013yoa}, and the Starobinsky model \cite{Starobinsky:1980te}, in which the role of the inflaton is played by a scalar degree of freedom hidden in the gravitational sector itself; an encyclopedic survey of single-field models can be found in \cite{Martin:2013tda}. Many of these models fit the present observational data well. However, the identity of the inflaton in all of them is decoupled from the known particle physics: the inflaton mass, its couplings, and its interactions with the SM fields (and thus the physics of reheating) are largely free parameters. While such extensions are certainly possible, they come at the cost of enlarging the particle content of the theory without direct experimental evidence.

The discovery of the Higgs boson at the Large Hadron Collider in 2012 completed the field content of the SM and established the existence of an elementary (no signs of compositeness are seen in experiments) scalar field in Nature.  It is the only fundamental scalar known to exist.  It is therefore natural to ask whether the Higgs field itself could play the role of the inflaton. Higgs inflation \cite{Bezrukov:2007ep} provides an affirmative answer to this question.

In the previous chapters of this article, we considered relatively small values of the Higgs field, of the order of the Fermi scale. Also, gravity did not play an essential role. Now, to determine the very early Universe evolution in the Standard Model, we should add gravity and specify the interactions of the SM fields with the gravitational degrees of freedom. 

There is no unique definition of gravitational degrees of freedom: in {\em metric gravity}, it is the metric which determines both distances and the parallel transport of the vectors; in {\em Palatini gravity},  the metric and connection are treated as independent variables; in {\em Einstein--Cartan gravity} these are the tetrads and Lorentz gauge fields -- comprising the spin connection and resulting in the appearance of torsion. Even more complicated objects appear if the gravitational theory contains non-metricity. These different possibilities give rise to distinct interactions of the Higgs field with gravity, discussed in what follows. 

\subsection{Higgs Inflation in Metric Gravity}
\label{metricHI}
Following the ideas of induced gravity \cite{Sakharov:1967pk}, as a first step, one can consider the SM on a curved background, taking the metric as a classical field. To get the diffeomorphism invariant action, change the integration measure by the invariant one, $d^4x \to  \sqrt{g}d^4x$  (here $g$ is the determinant of the metric with the minus sign) and replace the derivatives by curved space covariant derivatives. The quantum loops of the SM fields will generate an infinite number of diffeomorphism-invariant terms in the effective Lagrangian, dependent on the metric. There are several structures with divergent coefficients, which can be classified by the dimensionality of the operators.

The lowest order operator, the cosmological constant $\Lambda$,  is independent of the metric and the SM fields. The next one is a term linear in scalar curvature - Ricci scalar $R$ (dimensionality GeV$^2$). They give nothing but the  Einstein-Hilbert action 
\be
\label{EHact}
L_{\rm{EH}} = -\frac{M_P^2}{2} \int \diff^4 x \sqrt{g}(L_0+L_2)\,, ~~L_0=\Lambda\,,~~L_2=R\,.
\ee

 The next-order pure metric terms represent operators of mass dimension four. These are:
\begin{equation}
\label{L4}
    L_4= \frac{1}{4f_R^2} R^2+C_1 W_{\mu\nu\lambda\rho}W^{\mu\nu\lambda\rho}+ C_2 E_4 +C_3 \Box R\,.
\end{equation}
Here $\Box=g^{\mu\nu}\nabla_\mu \nabla_\nu$ is the covariant D'Alembertian, and $f_R^2,~C_{1,2,3}$ are dimensionless constants, while
\begin{equation}
    E_4=W_{\mu\nu\lambda\rho}W^{\mu\nu\lambda\rho}+\frac{2}{3}R^2-2 R_{\mu\nu}R^{\mu\nu}\,,
\end{equation}
is the Euler density, and $W_{\mu\nu\lambda\rho}$ the Weyl tensor.  The terms $E_4$ and $ \Box R$ are total derivatives.  In the presence of matter,  there is also the non-minimal coupling of the Higgs field to the Ricci scalar
\be
\label{nonmin}
L_{\xi}=-\xi_h H^\dagger H R\,.
\ee

As all parameters receive infinite contributions from the loop diagrams,  the coefficients $M_P, \Lambda,f_R^2,~C_{1,2,3}$ and $\xi_h$ cannot be computed within the SM (see Section \ref{Naturalness}) and should be taken from experiment, at least with our present level of knowledge.

The next step is to promote the metric to a dynamical field and decide which terms in equations (\ref{EHact}, \ref{L4}, \ref{nonmin}) we need to keep to describe its evolution. The full derivatives $E_4$ and $\Box R$ can be dropped (they do not contribute to the equations of motion). The term proportional to $C_1$  leads to the presence of extra degrees of freedom, some of them with the ``wrong'' sign of kinetic terms, resulting in instabilities and the presence of ``ghost'' particles, arguably leading to an inconsistent theory.\footnote{The presence of new degrees of freedom can make the theory renormalisable \cite{Stelle:1976gc}. Several articles (see, e.g.  \cite{Holdom:2024onr, Salvio:2015gsi}) argue that the ghost fields may be harmless.} So, to tread lightly, it is better to choose $C_1=0$. 

The unique theory, including operators up to mass-dimension four, which is free of ghosts, is that with a Lagrangian
\be
\label{Sact}
L_S=L_{\rm{EH}} + L_{\rm{SM~or~\nuMSM}}+ \frac{1}{4f_R^2} R^2-\xi_h H^\dagger H R\,. 
\ee
In addition to the graviton and the fields of the SM or \nuMSM, this theory contains just one extra spin-zero degree of freedom - the ``scalaron''. The simplest way to see that is to introduce an auxiliary field $\sigma$ (with mass-dimension one) that allows us to make the gravitational part of the action linear in the scalar curvature:
\be
\label{Sacteq}
 \frac{1}{4f_R^2} R^2 \to \frac{1}{2} \left(R \sigma^2 -\frac 1 2 f_R^2 \sigma^4\right) \,.
\ee
The equations of motion for $\sigma$ lead to $\sigma^2=R/f_R^2$, and inserting this solution into (\ref{Sacteq}) gives  (\ref{Sact}).  If the term $\propto R^2$ is absent in the action (corresponding to the choice $\frac{1}{4f_R^2} =0$), the theory only has the SM (or \nuMSM) degrees of freedom and the graviton.

The theory  (\ref{Sact}) is very interesting.  Depending on the relation between the parameters $\xi_h$ and $f_R^2$,  it embraces two successful inflationary models, consistent with all current cosmological observations  -  that of Starobinsky \cite{Starobinsky:1980te} (for $\xi_h< f_R \sim 10^5$) and Higgs inflation \cite{Bezrukov:2007ep} (for $10^5 \sim \xi_h\gtrsim f_R$).

Below, we will only consider Higgs inflation and thus omit the $R^2$ term from the action (\ref{Sacteq}). Then the theory is minimal in the sense that it only contains known degrees of freedom and has just one extra new parameter - the strength of the non-minimal coupling $\xi_h$,  which must be fixed by cosmological observations. The importance of the non-minimal coupling of the inflaton to gravity was understood already in 
\cite{Spokoiny:1984bd, Salopek:1988qh, Kaiser:1994vs}. 

The Higgs-gravity action has the form
\begin{equation}
\label{HIact}
S = \int d^4x \sqrt{-g}\left[
-\frac{M_P^2+\xi_h h^2}{2}R +\frac12 (\partial_\mu h)^2 -\frac{\lambda}{4}(h^2-v^2)^2
\right],
\end{equation}
written here in the unitary gauge, $h^2=H^\dagger H/2$.  For large field values, $\xi_h h^2 \gg M_P^2$, the effective Planck mass scales as $\propto h^2$, the (zero temperature) vev of the Higgs field $v$ can be neglected, and the theory acquires an approximate scale invariance.  This is precisely the regime relevant for inflation.

\begin{figure}[tbh]
\centerline{\includegraphics[width=7cm]{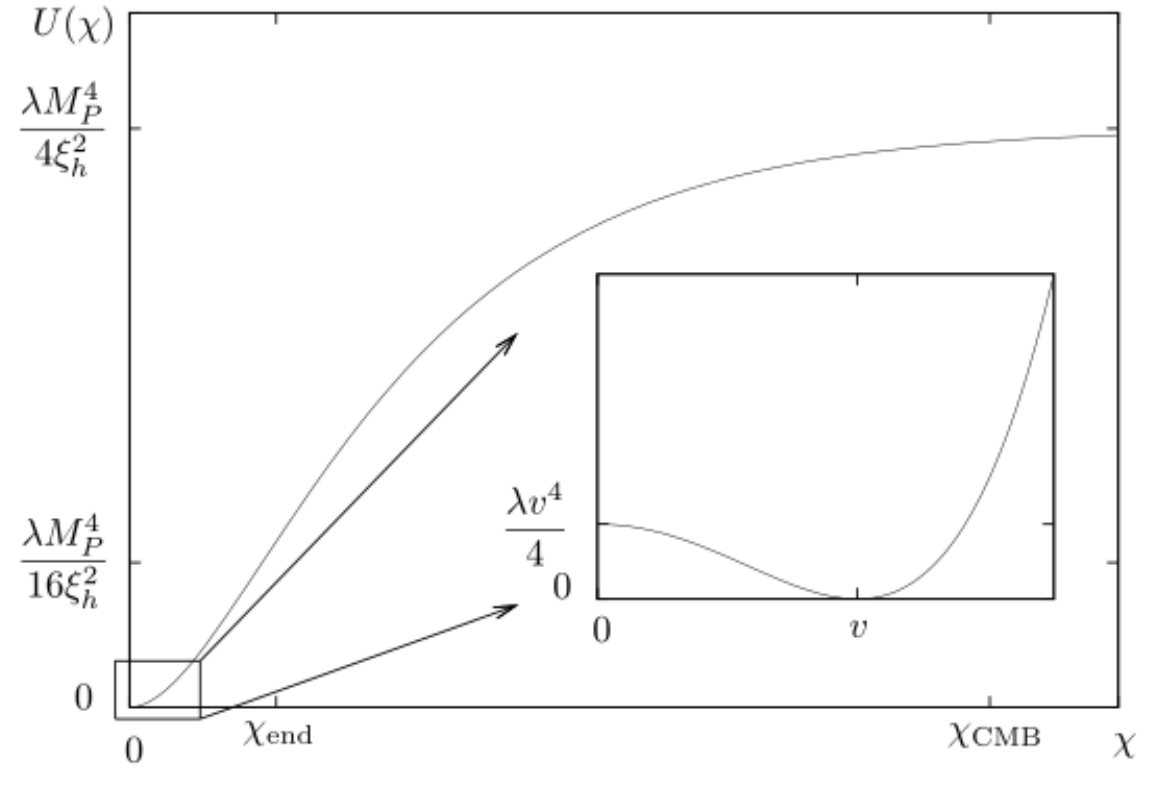} }
\caption{Effective potential in the Einstein frame.  The insert magnification is not to scale.} 
\label{Higgs_pot}
\end{figure}

It is convenient to perform a Weyl rescaling to the Einstein frame, in which the gravitational sector takes the canonical Einstein-Hilbert form $-M_P^2/2 R$:
\begin{equation}
\label{conf}
  g_{\mu\nu}\to  \hat{g}_{\mu\nu} = \Omega^2 g_{\mu\nu}
  \;,\quad
  \Omega^2 = \frac{M_P^2+\xi_h h^2}{M_P^2}\,.
\end{equation}
This transformation leads to a non-canonical kinetic term for the Higgs field,
\be
\label{KT}
 \frac{1}{2}K(h) (\partial_\mu h)^2 \,,
\ee
where
\be
\label{KTT}
K(h)=\dfrac{1}{\O^2}+\dfrac{6\xi_h^2h^2}{M_P^2\O^4}\,,
\ee
and to a modified potential 
\be
\label{Vmod}
\frac{\lambda}{4\Omega^4}(h^2-v^2)^2\,.
\ee
After an appropriate field redefinition to a canonically normalised scalar $\chi$, one finds that the scalar potential for $\chi$ has an exponentially flat plateau at large field values (see Fig. \ref{Higgs_pot})
\begin{equation}
\label{flat}
U(\chi)\simeq \frac{\lambda M_P^4}{4\xi_h^2} \left(1-e^{-\sqrt{2/3}\chi/M_P}\right)^2 .
\end{equation}
This flattening of the potential is the key mechanism underlying Higgs inflation.

Inflation proceeds in the standard slow-roll regime. Starting from large field values $h \gg M_P/\sqrt{\xi_h}$, the Higgs field slowly rolls along the plateau, driving an extended period of quasi-de Sitter expansion. The value $\chi_{\rm CMB}$ in Fig. \ref{Higgs_pot} corresponds to the COBE Cosmic Microwave Background (CMB) scale entering the horizon.  Inflation ends when the slow-roll conditions break down at $h \sim M_P/\sqrt{\xi_h}$, after which the Higgs field drifts towards the electroweak minimum and transfers its energy into Standard Model particles ($\chi_{\rm end}$ in Fig. \ref{Higgs_pot} ). This is a computable process, since the couplings of the Higgs boson to the SM particles are known.  As the gauge couplings in the SM are of the order of one, this automatically guarantees efficient reheating after inflation \cite{Bezrukov:2007ep, Bezrukov:2008ut, Garcia-Bellido:2008ycs}. The reheating temperature is $T_{\rm reh} \simeq 4 \times 10^{15}$ GeV \cite{DeCross:2015uza, DeCross:2016fdz,  DeCross:2016cbs, Ema:2016dny, Sfakianakis:2018lzf}. Thus, the Higgs field not only drives inflation but also reheats the Universe and serves as the origin of the hot Big Bang.

Matching the amplitude of scalar perturbations fixes the non-minimal coupling to be approximately
\begin{equation}
\xi_h \simeq 5\times 10^4 \sqrt{\lambda},
\end{equation}
giving $\xi_h \simeq 5\times 10^3$ if we take the scalar self-coupling of the Higgs field $\lambda \simeq 10^{-2}$ at high energy scale, see Fig. \ref{lambda}.

The inflationary predictions are in agreement with present observations (some tension appears in the (potentially problematic) combination of the data of \cite{AtacamaCosmologyTelescope:2025nti, AtacamaCosmologyTelescope:2025blo, DESI:2024uvr, DESI:2024mwx}, giving $n_s = 0.9743 \pm 0.0034$). To leading order in the number of e-folds $N$, one finds
\begin{equation}
\label{inN}
n_s \simeq 1-\frac{2}{N},
\qquad
r \simeq \frac{12}{N^2}.
\end{equation}
For $N\simeq 60$, this gives \cite{Bezrukov:2007ep,Bezrukov:2017dyv}
\begin{equation}
\label{num}
n_s \simeq 0.966,
\qquad
r \simeq 0.0033.
\end{equation}

The discussion of the inflationary behaviour above was based on the classical field theory (\ref{flat}). Clearly, the quantum consistency of the theory should be analysed. Since gravity is non-renormalisable, Higgs inflation must be understood as an effective field theory valid below an appropriate cutoff scale. There are numerous articles (see, e.g. \cite{Barbon:2009ya, Burgess:2010zq, Bezrukov:2010jz, Giudice:2010ka, Bauer:2010jg, Rasanen:2017ivk, Shaposhnikov:2020fdv, Langvik:2020nrs} ) analysing quantum aspects of Higgs inflation (see also a recent review \cite{Shaposhnikov:2025eyp}). The list of questions includes: 
\begin{enumerate}
\item What is the domain of energies and of the Higgs field background in which the perturbative analysis is possible?
\item Does inflation take place in this region?
\item What are the quantum corrections to inflationary observables?
\item What is the running of the SM effective potential in a theory with non-minimal coupling from the Fermi to the inflationary energy scale?
\item Can we have Higgs inflation if the Higgs scalar self-coupling changes its sign at the energy scale $\sim 10^{10}$ GeV (metastable vacuum situation), i.e. well below the inflationary scale $M_P/\sqrt{\xi_h} \sim 10^{16}$ GeV?
\end{enumerate}
And, of course, a more fundamental question: `` Do we need any new degrees of freedom in addition to the graviton and those of the SM/\nuMSM to make the theory self-consistent quantum-mechanically at any energy scale?'' 

While the answer to the last question remains unknown, the effective field theory for Higgs inflation constructed in \cite{Bezrukov:2010jz} and developed in \cite{Bezrukov:2014ipa} allows us to answer the others. In short, inflation occurs in a weakly-coupled domain of the effective field theory, making the classical predictions of the Higgs inflation reliable. The reheating of the Universe, though, occurs in a regime which goes beyond weak coupling and thus cannot be described reliably \cite{DeCross:2015uza, DeCross:2016fdz, DeCross:2016cbs, Ema:2016dny, Sfakianakis:2018lzf, Bezrukov:2010jz}, but the conclusion about the high value of the reheating temperature remains. The renormalisation group running of the effective action to large Higgs fields can be done with the {\em standard} $\beta$-functions of the SM up to the Higgs field values $h_*\simeq M_P/\xi_h$, which is smaller than the inflationary scale. The latter fact introduces uncertainties in the mapping between the low-energy couplings and the couplings at the inflationary scale due to threshold effects at  $h_*$. They, however, do not influence the inflationary predictions (\ref{inN},\ref{num}) \cite{Bezrukov:2010jz}.  A metastable vacuum does not exclude Higgs inflation \cite{Bezrukov:2014ipa}. The viability of the scenario depends on the detailed shape of the RG-improved effective potential and on threshold corrections relating the low-energy Standard Model to the inflationary regime. In particular, finite matching corrections at  $h \simeq h_*$ can restore a suitable inflationary plateau $h \gtrsim M_P/\sqrt{\xi_h}$  even when the low-energy vacuum is metastable, whereas reheating of the Universe after inflation restores the symmetry and brings the system eventually to the vacuum at the Fermi scale. (This is the point anticipated in Section \ref{consistency}: the fate of a metastable EW vacuum, harmless for the late Universe, becomes a non-trivial quantitative question for the earliest one.)

It is worth summarising the characteristic consequences of the hypothesis that the inflaton is the Higgs boson, which distinguish it from the models with a new inflaton field. First, no new particle degrees of freedom are introduced: the metric gravity scenario is maximally economical, and its single new parameter $\xi_h$ is fixed by the amplitude of the observed CMB perturbations. Second, since the couplings of the Higgs boson to all the SM particles are experimentally known, the physics of reheating is fixed: the reheating temperature is computable, and hence so is the number of e-folds $N$ between the horizon exit of the CMB scales and the end of inflation.  As a result, $n_s$ and $r$ are predictions (\ref{num}) rather than functions of free parameters. 
Third, the scenario may correlate collider and cosmological observables up to uncertainties at the energy scale $h_*$: the same combination of the Higgs mass, the top quark mass, and the strong coupling constant that determines the running of $\lambda$ (Section \ref{consistency}) also controls the viability of inflation and the radiative corrections to it, so that improved measurements of $m_t$ at colliders bear directly on the physics of the early Universe. Finally, as discussed in the following subsections, the value of the tensor-to-scalar ratio $r$ discriminates between the gravitational formulations of the same particle-physics theory, making the search for primordial gravitational waves a probe of both the inflaton identity and the structure of gravity.

\subsection{Beyond Metric Gravity}
\label{EC}
The Lagrangian (\ref{HIact}) is based on the metric formulation of gravity, where the role of the dynamical variable is played by $g_{\mu\nu}$.  This formulation is not unique. For example, in Palatini gravity \cite{Palatini:1919ffw},  the metric and connection are treated as independent variables. Moreover, the metric setup is not even a natural one when we need to describe fermions, which are an essential part of the SM. A curved-space action for fermions, in addition to diffeomorphisms (Diff), is invariant under local Lorentz transformations. Thus, the fermion Lagrangian contains new (in comparison with the metric) geometric objects: the tetrad - $e^A_\mu$ and the gauge field of the local Lorentz group, the spin connection $\omega^{AB}_\mu$ (for a review, see ~\cite{Hehl:1976kj}). These variables are a priori independent (the tetrad and spin connection totalise 40 components, 16 and 24 respectively) and serve as a basis of the Einstein-Cartan (EC) formulation of gravity \cite{Cartan:1922, Cartan:1923, Cartan:1924, Cartan:1925}. The EC formulation fits well with the gauge principle used in the construction of the SM: to get the SM mediators of the strong and electroweak interactions, the global internal SU(3)xSU(2)xU(1) symmetry is gauged, whereas to get gravity, the global Lorentz symmetry is gauged ~\cite{Utiyama:1956sy, Kibble:1961ba, Sciama:1962}.  In pure gravity, all these formulations reproduce classical General Relativity. However, once scalar and fermion fields are coupled non-minimally to curvature, their dynamics differ, and the resulting cosmological scenarios are no longer equivalent \cite{Shaposhnikov:2020aen, Langvik:2020nrs, Shaposhnikov:2020gts, Karananas:2021zkl}.

As an example, let us consider EC gravity. In the EC formulation, the tetrad plays the role of the ``gauge field'' of translations and the spin connection that of the local Lorentz group, in full analogy with the SM gauge fields. Each has its own field strength: for the spin connection this is the curvature, and for the tetrad it is the torsion. The crucial physical point, elaborated below, is that in the theories considered here torsion does not propagate: it enters the action without derivatives and can be integrated out algebraically. Doing so leaves the graviton and the SM fields as the only propagating degrees of freedom, but generates new contact interactions among the SM fields and modifies the kinetic term and the potential of the Higgs field. The equations of this subsection make this statement precise. The corresponding field strengths are the curvature
\be
\label{eq:curv_def}
F_{\m\n}^{AB} = \p_\m \omega_\n^{AB} -\p_\n \omega_\m^{AB}+\omega^A_{\m C}\omega^{CB}_\n - \omega^A_{\n C}\omega^{CB}_\m \ , 
\ee
and torsion
\be
\label{eq:tors_def}
T_{\m\n\rho}=e_{\mu A} T^A_{\n\rho}\ ,~~~T_{\m\n}^A = \p_\m e_\n^A -\p_\n e_\m^A+ \omega^A _{\m B}e^B_\n -\omega^A _{\n B}e^B_\m  \,. 
\ee
Torsion can be decomposed into irreducible vector  ($v_\mu,~a_\mu$) and tensor ($\tau_{\m\n\rho}$) components as:
\be 
\label{eq:tors_all}
v_\m = T^\n_{~\m\n} \ ,~~~a^\m = E^{\m\n\rho\sigma}T_{\n\rho\sigma} \ ,~~~\tau_{\m\n\rho} =\frac 2 3 \l( T_{\m\n\rho} -v_{[\n} g_{\rho]\m} - T_{[\n\rho]\m} \r) \,,
\ee
where the square brackets mean antisymmetrisation with respect to the pair of indices, and $ E^{\m\n\rho\sigma}=\epsilon^{\m\n\rho\sigma}/det(e)$. Two scalar curvature invariants read
\be
\label{eq:f_def}
F \equiv \frac {1}{8} \epsilon_{ABCD}E^{\m\n\rho\sigma}F^{AB}_{\m\n}e^C_\rho e^D_\sigma \ ,~~~\text{and}~~~\tilde F \equiv E^{\m\n\rho\sigma}e_{\rho C}e_{\sigma D}F_{\m\n}^{CD}  \ .
\ee
The second structure is called the Holst term.

To find a relation between the EC formulation of gravity and the metric one, the following formulas are helpful:
\begin{align}
\label{eq:f_decomp}
F & = \frac{R}{2} +\frac{1}{\sqrt{g}}\p_\m\l\sqrt{g}v^\m\r) -\frac 1 3 v_\m v^\m + \frac{1}{48} a_\m a^\m +\frac 1 4 \tau_{\m\n\rho}\tau^{\m\n\rho}\ ,  \\
\label{eq:tilde_f_decomp}
\tilde F & = -\frac{1}{\sqrt{g}}\p_\m\l\sqrt{g}
a^\m\r)+\frac 2 3 a_\m v^\m -\frac 1 2 \epsilon^{\m\n\rho\sigma}\tau_{\lambda\m\n}\tau^{\lambda}_{~\rho\sigma}\ , 
\end{align}
where the torsion-free Riemannian curvature (and the Ricci scalar $R$) is defined as usual in metric gravity.

The analogue of (\ref{EHact}) in EC gravity, involving operators of dimension up to two, is
\begin{align}
\label{EC2}
S_{\rm{EC},2} = - M_P^2  \int \diff^4 x \det{e} [F +&a  \tilde F+b v_\mu v^\mu +c a_\mu a^\mu + d a_\mu v^\mu  +\\
 &e \tau^{\mu\nu\rho} \tau_{\mu\nu\rho}+f E^{\mu\nu\rho\sigma}\tau^\lambda_{\mu\nu}\tau_{\lambda\rho\sigma}+\Lambda]\,,
\nonumber
\end{align}
where $a,b,...,f$ are dimensionless constants.  Despite the presence of many variables, this action is equivalent to (\ref{EHact}) and contains just a graviton. Indeed, with the use of (\ref{eq:f_decomp},\ref{eq:tilde_f_decomp}), one can see that the equations of motion for the torsion give $v_\mu=a_\mu=\tau_{\m\n\rho}=0$, leaving us with the Einstein-Hilbert action. 

This situation changes when the SM is added. Now, non-minimal couplings of $H^\dagger H$ to all the structures appearing in (\ref{EC2}) are allowed. In the resulting theory, torsion remains non-dynamical; though the vectors $v_\mu$ and $a_\mu$ are not equal to zero anymore, and are expressed through the SM fields via the equations of motion. Thus, the only propagating degrees of freedom are those of the SM and the graviton. Instead of just one extra parameter - $\xi_h$, the theory depends now on the plethora of non-minimal couplings, leading to a kinetic terms and potential for the Higgs different from those in eq. (\ref{KTT},\ref{Vmod}).  With a specific choice of non-minimal couplings, the EC gravity can be converted \cite{Shaposhnikov:2020gts} to the metric gravity discussed above, or the Palatini gravity \cite{Palatini:1919ffw}. 

The Palatini case is interesting for several reasons \cite{Bauer:2008zj}. The action is still given by eq. (\ref{HIact}), but now the metric and connection are treated as independent variables. The connection is non-dynamical and is expressed through the metric and the Higgs field via equations of motion; torsion is absent. The physical difference from the metric case is easy to state. In metric gravity, the Weyl rescaling (\ref{conf}) converts the $h$-dependence of the effective Planck mass into an extra contribution to the Higgs kinetic term -- the second term in eq. (\ref{KTT}) -- because the connection, being built from the metric, feels the rescaling. In Palatini gravity the connection is an independent variable, and this contribution is absent: the canonically normalised inflaton is a different function of $h$, the inflationary plateau starts at larger $\xi_h$ and has a lower energy scale. Since the potential energy at the end of inflation is smaller, while the Higgs couplings to the SM particles are the same, the resulting reheating temperature is lower than in the metric case, and the amplitude of tensor perturbations is strongly suppressed. In the Einstein frame, the kinetic term for the Higgs field differs from (\ref{KTT}) and reads
\be
K(h)=\dfrac{1}{\O^2}\,,
\label{palat}
\ee
whereas the potential is still given by (\ref{Vmod}).

The predictions of Palatini-Higgs inflation are different from those of the metric theory. The relation between the non-minimal coupling and scalar self-coupling now reads \cite{Bauer:2008zj}
\be
\xi_h=2\times 10^{10}\lambda\,,
\ee
meaning that the value of $\xi$ in the Palatini case should be substantially larger than in the metric case, $\xi_h \simeq 2\times 10^8$ for $\lambda \simeq 10^{-2}$.  Because of that, the reheating temperature is about two orders of magnitude smaller, $T_{\mathrm{reh}}\simeq\unit[2\times10^{13}]{GeV}$. The number of e-foldings is somewhat smaller than in the metric case, $N\simeq 55$. The spectral tilt is still given by (\ref{num}), but the tensor-to-scalar ratio is predicted to be
\be
r=\dfrac{2}{\xi_h N^2}\,,
\ee
i.e. much smaller than in the metric case. Because of the modification of the kinetic term (\ref{palat}), the theory remains in the perturbative domain all the way from inflation to low energies \cite{Bauer:2010jg, Rasanen:2017ivk, Shaposhnikov:2020fdv}.

A more general set of parameters allows for nontrivial but nondynamical torsion, making it possible to smoothly interpolate between different possibilities. A robust conclusion of this generalised perspective is that Higgs inflation is not an isolated mechanism but a broad and generic phenomenon \cite{Shaposhnikov:2020gts, Langvik:2020nrs}. Across large regions of parameter space, successful slow-roll inflation occurs naturally. The scalar spectral index $n_s$ is remarkably stable and remains close to the universal attractor value given by (\ref{inN}),  with only mild dependence on the precise gravitational formulation; deviations from this number are also possible for a specific choice of the parameters.

By contrast, the tensor-to-scalar ratio is highly sensitive to the geometry of the gravitational sector. In the metric formulation, one finds the familiar prediction $r\simeq 0.003$, while in Palatini and more general Einstein--Cartan realisations, the value of $r$ may be strongly suppressed or enhanced, ranging over many orders of magnitude \cite{Langvik:2020nrs, Shaposhnikov:2020gts}. This sensitivity makes primordial tensor modes a particularly powerful observational discriminator among different realisations of Higgs inflation. A future detection of primordial gravitational waves would therefore provide not only information about inflation, but potentially also about the geometric structure of gravity itself.

\subsection{Weyl-invariant EC gravity} 
\label{Weyl}
The EC gravity provides a natural realisation \cite{Obukhov:1982zn, Dereli:1982xb} of the Weyl symmetry, which can considerably constrain the structure of the theory. Weyl symmetry changes the length of a ruler locally, and the corresponding theory does not have any intrinsic mass parameters; all coupling constants are dimensionless. In the flat space limit, Weyl symmetry leads to a conformal field theory (CFT). Interestingly, the SM and the \nuMSM are invariant under the full conformal group if the mass of the Higgs boson and Majorana masses of HNLs are set to zero, whereas the energy of the ground state in CFTs is equal to zero, hinting towards understanding of the tiny values of the Higgs and HNL masses, and the cosmological constant in comparison with the Planck scale.  The spontaneous breakdown of the Weyl and conformal symmetries brings mass scales into existence. 

The formulas below serve a single purpose: to exhibit how the geometric variables of EC gravity respond to a local change of units, and to write down the most general action compatible with this symmetry. The key observation is that one of the components of torsion -- the vector $v_\mu$ -- transforms exactly as an abelian gauge field of the Weyl symmetry. It is thus the geometry itself, with no extra fields added by hand, that provides the ``compensator'' needed to render the theory Weyl-invariant. Under the Weyl transformation, the tetrad, spin connection, and torsion transform as follows \cite{Obukhov:1982zn, Dereli:1982xb}:
\ba
\label{WT}
e_\mu^A\to Q(x) e_\mu^A,~~~ \omega_\mu^{AB} \to \omega_\mu^{AB}\,
a_\mu\to a_\mu, ~~v_\mu\to v_\mu +3Q(x)\partial_\mu Q(x)^{-1},~~ \tau_{\mu\nu\rho}\to Q(x)^2\tau_{\mu\nu\rho}\,,
\ea
where $Q(x)$ is an arbitrary function of space-time coordinates. The vector torsion $v_\mu$ transforms as a (non-compact) U(1) gauge field, allowing us to write a Weyl-covariant derivative and thus convert any action into the Weyl-invariant one. The Weyl-invariant ghost-free action $S_{ECWI}$  has the form
\be
\label{ECWI}
S_{ECWI}=S_{\rm{EC},4} +S_{\rm{HWI}}\,,
\ee
where the gravity part is 
\be
\label{EC4}
S_{\rm{EC},4}= \int \diff^4 x \det{e} \left[\f{1}{f^2} F^2 + \f{1}{\tilde f^2} \tilde F^2 + \f{2}{f_m^2} F \tilde F\right]\,,
\ee
and  $f,~\tilde f$ and $f_m$ can be called the local Lorentz group gauge couplings.
The Higgs part is 
\be
\label{HWI}
S_{\rm{HWI}}
=  \f{1}{2} (D_\mu h)^2-\f{\lambda}{4}h^4 - h^2\left(\xi_h  F +\zeta_h  \tilde F+ c_a a_\mu^2\right)
\ee
with the Weyl-covariant derivative
\be
\label{weylder}
D_\mu = \partial_\mu+\f{1}{3}v_\mu\,.
\ee
 As previously, $h$ is the Higgs field in unitary gauge (we did not include in (\ref{HWI}) the terms proportional to the squares of the tensor torsion and the Higgs field as they are nullified by the equations of motion) .

This theory, in addition to the SM (or $\nu$MSM) degrees of freedom and a graviton, contains one extra spinless particle corresponding to a scalar field, say $a$, hidden in the gravitational degrees of freedom. Since the Weyl symmetry is local, there is no massless Goldstone mode. The spectrum is massive, with the exception of the graviton and the photon. 

This theory allows us to unify the strong CP and hierarchy puzzles \cite{Karananas:2024xja}. If the couplings $f,~\tilde f$, and $f_m$  are tiny and of the same order of magnitude, the tree, classical masses of the Higgs boson, the field $a$, and the cosmological constant are also small, as all these quantities are proportional to them. For simplicity of presentation we set $f_m=0$; in this case the minimum of the combined Higgs - $a$ potential is at
\be
\label{min}
h^2=\frac{f^2 \xi_h M_P^2}{4\lambda},~~a^2=0\,,
\ee
and the masses of the excitations are
\be
\label{masses}
m_h^2=- \frac{f^2  \xi_h M_P^2}{4}\,,~~
m_a^2= \frac{M_P^2 \tilde{f}^2}{48} \,.
\ee
The vacuum energy of the system is
\be
\label{evac}
\epsilon_{vac}\approx\frac{f^2  M_P^4}{16}\,.
\ee
Switching on $f_m\neq 0$ does not change this picture qualitatively: the minimum of the potential is shifted, the excitations of $h$ and $a$ mix, and the masses and the vacuum energy remain parametrically small, being proportional to the small couplings $f,~\tilde f,~f_m$. It is this general case, with all three couplings present, that is used in \cite{Karananas:2024xja} for the unified treatment of the strong CP and hierarchy problems.

Taken at face value, equations (\ref{min}-\ref{evac}), taken together,  contradict observations. If we fix, for instance, $f$, fitting Eq. (\ref{evac}) to the observed value of the cosmological constant, the Higgs boson mass, following from Eq. (\ref{masses}), will be way too small in comparison with experiment. If this smallness is compensated by a huge value of $\xi_h$, the theory will contradict the known Higgs boson properties \cite{Atkins:2012yn}. These tree formulas, however,  receive both perturbative and non-perturbative corrections. The extension of the classical theory to the quantum one can be done in a Weyl-invariant way \cite{Englert:1976ep, Wetterich:1987fm, Shaposhnikov:2008xi, Ghilencea:2015mza, Ghilencea:2016dsl, Ghilencea:2017yqv,  Ghilencea:2016ckm, Ferreira:2018itt, Ghilencea:2024usf, Shaposhnikov:2025znm}, ensuring the stability of (\ref{masses}) in perturbation theory. In particular, the values $f=0$ and $\tilde{f}=0$ are fixed points of the RG evolution. A possible coupling of the fields $a$ and $h$ to the QCD topological invariant $G_{\m\n}^a G^{\m\n}_a$ will result in a QCD contribution to their mass. The gravity-Higgs instantons, discussed in \cite{Shaposhnikov:2018xkv, Shaposhnikov:2018jag, Shaposhnikov:2020geh}  may lead to $m_h \propto M_P \exp{(-S)}$ (here $S$ is the instanton action), curing the mismatch between (\ref{min},\ref{evac},\ref{masses}) and experiment, if $S\sim 37$. 

The field $a$ -- an axion-like particle (ALP)  has a non-trivial coupling to the axial and vector fermionic currents, and has all the requisites to solve the strong CP-problem, provided its ``gravitational'' mass is small compared with the QCD induced mass \cite{Karananas:2024xja, Karananas:2025ews}. It has a very distinct cosmology and phenomenology \cite{Karananas:2025uhy} from the standard axion associated with the breaking of the Peccei-Quinn symmetry \cite{Peccei:1977hh}. The theory naturally accommodates metric Higgs inflation \cite{Karananas:2025qsm}.

\section{Conclusions}
\label{Conclusions}
In this paper, we considered two minimal models of particle physics: the Standard Model and the \nuMSM. Both are self-consistent quantum field theories up to very high energies, presumably up to the Planck scale. We explained why the ``naturalness'' arguments, calling for the appearance of new physics right above the Fermi scale, rest on an artefact of the computational procedure rather than on a property of the theory. Therefore, these models provide a self-consistent framework for describing the evolution of the Universe from inflation to the present epoch.

We described the possible roles of the Higgs field in cosmology. In particular, 
\begin{itemize}
\item the Higgs boson could inflate the Universe and make it  flat, homogeneous and isotropic;
\item quantum fluctuations of the Higgs field could lead to structure formation;
\item after cosmological inflation, the Higgs field could initiate the Hot Big Bang and produce all matter in the Universe;
\item the Yukawa couplings of the Higgs boson in the \nuMSM could play a crucial role in baryogenesis, leading to a charge asymmetric Universe;
\item dark matter production may arise from mixing between neutrinos and Heavy Neutral Leptons, induced by the Higgs field. 
\end{itemize}
Thus, the Higgs boson of the SM, in addition to breaking the electroweak symmetry and generating the masses of quarks, leptons, and intermediate vector bosons, is a crucial element in understanding the structure of the Universe.

\section*{Acknowledgements}
I thank my collaborators Takehiko Asaka, Fedor Bezrukov, Steve Blanchet, Alexey Boyarsky, Dmitry Gorbunov, Kostas Farakos, Keijo Kajantie, Mikhail Kalmykov, Georgios Karananas, Sergei Khlebnikov, Bernd Kniehl, Vadim Kuzmin, Mikko Laine, Amaury Magnin, Valery Rubakov, Javier Rubio,  Oleg Ruchayskiy, Kari Rummukainen, Sergei Sibiryakov, Andrey Shkerin, Inar Timiryasov, and Sebastian Zell for developing together the ideas described in this review. I am especially grateful to Georgios Karananas for his comments on the manuscript.

\bibliographystyle{utphys} 
\bibliography{Review_Higgs.bib}

\providecommand{\href}[2]{#2}\begingroup\raggedright\begin{thebibliography}{100}

\bibitem{ParticleDataGroup:2024cfk}
{\bfseries Particle Data Group} Collaboration, S.~Navas {\em et~al.}, ``{Review
  of particle physics},''
  \href{http://dx.doi.org/10.1103/PhysRevD.110.030001}{{\em Phys. Rev. D}
  {\bfseries 110} no.~3, (2024) 030001}.

\bibitem{Maiani:1977cg}
L.~Maiani, G.~Parisi, and R.~Petronzio, ``{Bounds on the Number and Masses of
  Quarks and Leptons},''
  \href{http://dx.doi.org/10.1016/0550-3213(78)90018-4}{{\em Nucl. Phys. B}
  {\bfseries 136} (1978) 115--124}.

\bibitem{Cabibbo:1979ay}
N.~Cabibbo, L.~Maiani, G.~Parisi, and R.~Petronzio, ``{Bounds on the Fermions
  and Higgs Boson Masses in Grand Unified Theories},''
  \href{http://dx.doi.org/10.1016/0550-3213(79)90167-6}{{\em Nucl. Phys. B}
  {\bfseries 158} (1979) 295--305}.

\bibitem{Lindner:1985uk}
M.~Lindner, ``{Implications of Triviality for the Standard Model},''
  \href{http://dx.doi.org/10.1007/BF01479540}{{\em Z. Phys. C} {\bfseries 31}
  (1986) 295}.

\bibitem{Glashow:1961tr}
S.~L. Glashow, ``{Partial Symmetries of Weak Interactions},''
  \href{http://dx.doi.org/10.1016/0029-5582(61)90469-2}{{\em Nucl. Phys.}
  {\bfseries 22} (1961) 579--588}.

\bibitem{Weinberg:1967tq}
S.~Weinberg, ``{A Model of Leptons},''
  \href{http://dx.doi.org/10.1103/PhysRevLett.19.1264}{{\em Phys. Rev. Lett.}
  {\bfseries 19} (1967) 1264--1266}.

\bibitem{Salam:1968rm}
A.~Salam, ``{Weak and Electromagnetic Interactions},''
  \href{http://dx.doi.org/10.1142/9789812795915_0034}{{\em Conf. Proc. C}
  {\bfseries 680519} (1968) 367--377}.

\bibitem{Minkowski:1977sc}
P.~Minkowski, ``{$\mu \to e\gamma$ at a Rate of One Out of $10^{9}$ Muon
  Decays?},'' \href{http://dx.doi.org/10.1016/0370-2693(77)90435-X}{{\em Phys.
  Lett. B} {\bfseries 67} (1977) 421--428}.

\bibitem{Yanagida:1979as}
T.~Yanagida, ``{Horizontal gauge symmetry and masses of neutrinos},'' {\em
  Conf. Proc. C} {\bfseries 7902131} (1979) 95--99.

\bibitem{Gell-Mann:1979vob}
M.~Gell-Mann, P.~Ramond, and R.~Slansky, ``{Complex Spinors and Unified
  Theories},'' {\em Conf. Proc. C} {\bfseries 790927} (1979) 315--321,
  \href{http://arxiv.org/abs/1306.4669}{{\ttfamily arXiv:1306.4669 [hep-th]}}.

\bibitem{Glashow:1979nm}
S.~L. Glashow, ``{The Future of Elementary Particle Physics},''
  \href{http://dx.doi.org/10.1007/978-1-4684-7197-7_15}{{\em NATO Sci. Ser. B}
  {\bfseries 61} (1980) 687}.

\bibitem{Mohapatra:1979ia}
R.~N. Mohapatra and G.~Senjanovic, ``{Neutrino Mass and Spontaneous Parity
  Nonconservation},'' \href{http://dx.doi.org/10.1103/PhysRevLett.44.912}{{\em
  Phys. Rev. Lett.} {\bfseries 44} (1980) 912}.

\bibitem{Asaka:2005an}
T.~Asaka, S.~Blanchet, and M.~Shaposhnikov, ``{The nuMSM, dark matter and
  neutrino masses},''
  \href{http://dx.doi.org/10.1016/j.physletb.2005.09.070}{{\em Phys. Lett. B}
  {\bfseries 631} (2005) 151--156},
  \href{http://arxiv.org/abs/hep-ph/0503065}{{\ttfamily arXiv:hep-ph/0503065}}.

\bibitem{Strumia:2006db}
A.~Strumia and F.~Vissani, ``{Neutrino masses and mixings and...},''
  \href{http://arxiv.org/abs/hep-ph/0606054}{{\ttfamily arXiv:hep-ph/0606054}}.

\bibitem{Asaka:2005pn}
T.~Asaka and M.~Shaposhnikov, ``{The $\nu$MSM, dark matter and baryon asymmetry
  of the universe},''
  \href{http://dx.doi.org/10.1016/j.physletb.2005.06.020}{{\em Phys. Lett. B}
  {\bfseries 620} (2005) 17--26},
  \href{http://arxiv.org/abs/hep-ph/0505013}{{\ttfamily arXiv:hep-ph/0505013}}.

\bibitem{Schechter:1980gr}
J.~Schechter and J.~W.~F. Valle, ``{Neutrino Masses in SU(2) x U(1)
  Theories},'' \href{http://dx.doi.org/10.1103/PhysRevD.22.2227}{{\em Phys.
  Rev. D} {\bfseries 22} (1980) 2227}.

\bibitem{Landau:1954cxv}
L.~D. Landau, A.~A. Abrikosov, and I.~M. Khalatnikov, ``{An asymptotic
  expression for the photon Green function in quantum electrodynamics},''
  \href{http://dx.doi.org/10.1016/b978-0-08-010586-4.50085-7}{{\em Dokl. Akad.
  Nauk SSSR} {\bfseries 95} no.~6, (1954) 1177--1180}.

\bibitem{Bogolyubov:1959vck}
N.~N. Bogolyubov, A.~A. Logunov, and D.~V. Shirkov, ``{The Method of Dispersion
  Relations and Perturbation Theory},'' {\em Sov. Phys. JETP} {\bfseries 10}
  no.~3, (1960) 574--581.

\bibitem{Andreassen:2014eha}
A.~Andreassen, W.~Frost, and M.~D. Schwartz, ``{Consistent Use of Effective
  Potentials},'' \href{http://dx.doi.org/10.1103/PhysRevD.91.016009}{{\em Phys.
  Rev. D} {\bfseries 91} no.~1, (2015) 016009},
  \href{http://arxiv.org/abs/1408.0287}{{\ttfamily arXiv:1408.0287 [hep-ph]}}.

\bibitem{Altarelli:1994rb}
G.~Altarelli and G.~Isidori, ``{Lower limit on the Higgs mass in the standard
  model: An Update},''
  \href{http://dx.doi.org/10.1016/0370-2693(94)91458-3}{{\em Phys. Lett. B}
  {\bfseries 337} (1994) 141--144}.

\bibitem{Casas:1994qy}
J.~A. Casas, J.~R. Espinosa, and M.~Quiros, ``{Improved Higgs mass stability
  bound in the standard model and implications for supersymmetry},''
  \href{http://dx.doi.org/10.1016/0370-2693(94)01404-Z}{{\em Phys. Lett. B}
  {\bfseries 342} (1995) 171--179},
  \href{http://arxiv.org/abs/hep-ph/9409458}{{\ttfamily arXiv:hep-ph/9409458}}.

\bibitem{Casas:1996aq}
J.~A. Casas, J.~R. Espinosa, and M.~Quiros, ``{Standard model stability bounds
  for new physics within LHC reach},''
  \href{http://dx.doi.org/10.1016/0370-2693(96)00682-X}{{\em Phys. Lett. B}
  {\bfseries 382} (1996) 374--382},
  \href{http://arxiv.org/abs/hep-ph/9603227}{{\ttfamily arXiv:hep-ph/9603227}}.

\bibitem{Andreassen:2017rzq}
A.~Andreassen, W.~Frost, and M.~D. Schwartz, ``{Scale Invariant Instantons and
  the Complete Lifetime of the Standard Model},''
  \href{http://dx.doi.org/10.1103/PhysRevD.97.056006}{{\em Phys. Rev. D}
  {\bfseries 97} no.~5, (2018) 056006},
  \href{http://arxiv.org/abs/1707.08124}{{\ttfamily arXiv:1707.08124
  [hep-ph]}}.

\bibitem{Bezrukov:2012sa}
F.~Bezrukov, M.~Y. Kalmykov, B.~A. Kniehl, and M.~Shaposhnikov, ``{Higgs Boson
  Mass and New Physics},''
  \href{http://dx.doi.org/10.1007/JHEP10(2012)140}{{\em JHEP} {\bfseries 10}
  (2012) 140}, \href{http://arxiv.org/abs/1205.2893}{{\ttfamily arXiv:1205.2893
  [hep-ph]}}.

\bibitem{Degrassi:2012ry}
G.~Degrassi, S.~Di~Vita, J.~Elias-Miro, J.~R. Espinosa, G.~F. Giudice,
  G.~Isidori, and A.~Strumia, ``{Higgs mass and vacuum stability in the
  Standard Model at NNLO},''
  \href{http://dx.doi.org/10.1007/JHEP08(2012)098}{{\em JHEP} {\bfseries 08}
  (2012) 098}, \href{http://arxiv.org/abs/1205.6497}{{\ttfamily arXiv:1205.6497
  [hep-ph]}}.

\bibitem{Buttazzo:2013uya}
D.~Buttazzo, G.~Degrassi, P.~P. Giardino, G.~F. Giudice, F.~Sala, A.~Salvio,
  and A.~Strumia, ``{Investigating the near-criticality of the Higgs boson},''
  \href{http://dx.doi.org/10.1007/JHEP12(2013)089}{{\em JHEP} {\bfseries 12}
  (2013) 089}, \href{http://arxiv.org/abs/1307.3536}{{\ttfamily arXiv:1307.3536
  [hep-ph]}}.

\bibitem{Bednyakov:2015sca}
A.~V. Bednyakov, B.~A. Kniehl, A.~F. Pikelner, and O.~L. Veretin, ``{Stability
  of the Electroweak Vacuum: Gauge Independence and Advanced Precision},''
  \href{http://dx.doi.org/10.1103/PhysRevLett.115.201802}{{\em Phys. Rev.
  Lett.} {\bfseries 115} no.~20, (2015) 201802},
  \href{http://arxiv.org/abs/1507.08833}{{\ttfamily arXiv:1507.08833
  [hep-ph]}}.

\bibitem{Bezrukov:2014ina}
F.~Bezrukov and M.~Shaposhnikov, ``{Why should we care about the top quark
  Yukawa coupling?},'' \href{http://dx.doi.org/10.1134/S1063776115030152}{{\em
  J. Exp. Theor. Phys.} {\bfseries 120} (2015) 335--343},
  \href{http://arxiv.org/abs/1411.1923}{{\ttfamily arXiv:1411.1923 [hep-ph]}}.

\bibitem{Shaposhnikov:2009pv}
M.~Shaposhnikov and C.~Wetterich, ``{Asymptotic safety of gravity and the Higgs
  boson mass},'' \href{http://dx.doi.org/10.1016/j.physletb.2009.12.022}{{\em
  Phys. Lett. B} {\bfseries 683} (2010) 196--200},
  \href{http://arxiv.org/abs/0912.0208}{{\ttfamily arXiv:0912.0208 [hep-th]}}.

\bibitem{Branchina:2015nda}
V.~Branchina and E.~Messina, ``{Stability and UV completion of the Standard
  Model},'' \href{http://dx.doi.org/10.1209/0295-5075/117/61002}{{\em EPL}
  {\bfseries 117} no.~6, (2017) 61002},
  \href{http://arxiv.org/abs/1507.08812}{{\ttfamily arXiv:1507.08812
  [hep-ph]}}.

\bibitem{Branchina:2016bws}
V.~Branchina, E.~Messina, and D.~Zappala, ``{Impact of Gravity on Vacuum
  Stability},'' \href{http://dx.doi.org/10.1209/0295-5075/116/21001}{{\em EPL}
  {\bfseries 116} no.~2, (2016) 21001},
  \href{http://arxiv.org/abs/1601.06963}{{\ttfamily arXiv:1601.06963
  [hep-ph]}}.

\bibitem{Bentivegna:2017qry}
E.~Bentivegna, V.~Branchina, F.~Contino, and D.~Zappal{\`a}, ``{Impact of New
  Physics on the EW vacuum stability in a curved spacetime background},''
  \href{http://dx.doi.org/10.1007/JHEP12(2017)100}{{\em JHEP} {\bfseries 12}
  (2017) 100}, \href{http://arxiv.org/abs/1708.01138}{{\ttfamily
  arXiv:1708.01138 [hep-ph]}}.

\bibitem{Branchina:2019tyy}
V.~Branchina, E.~Bentivegna, F.~Contino, and D.~Zappal{\`a}, ``{Direct
  Higgs-gravity interaction and stability of our Universe},''
  \href{http://dx.doi.org/10.1103/PhysRevD.99.096029}{{\em Phys. Rev. D}
  {\bfseries 99} no.~9, (2019) 096029},
  \href{http://arxiv.org/abs/1905.02975}{{\ttfamily arXiv:1905.02975
  [hep-ph]}}.

\bibitem{Bogolyubov:1959bfo}
N.~N. Bogolyubov and D.~V. Shirkov, {\em {Introduction to the theory of
  quantized fields}}, vol.~3.
\newblock 1959.

\bibitem{Peskin:1995ev}
M.~E. Peskin and D.~V. Schroeder,
  \href{http://dx.doi.org/10.1201/9780429503559}{{\em {An Introduction to
  quantum field theory}}}.
\newblock Addison-Wesley, Reading, USA, 1995.

\bibitem{Georgi:1974sy}
H.~Georgi and S.~L. Glashow, ``{Unity of All Elementary Particle Forces},''
  \href{http://dx.doi.org/10.1103/PhysRevLett.32.438}{{\em Phys. Rev. Lett.}
  {\bfseries 32} (1974) 438--441}.

\bibitem{Georgi:1974yf}
H.~Georgi, H.~R. Quinn, and S.~Weinberg, ``{Hierarchy of Interactions in
  Unified Gauge Theories},''
  \href{http://dx.doi.org/10.1103/PhysRevLett.33.451}{{\em Phys. Rev. Lett.}
  {\bfseries 33} (1974) 451--454}.

\bibitem{Kim:1979if}
J.~E. Kim, ``{Weak Interaction Singlet and Strong CP Invariance},''
  \href{http://dx.doi.org/10.1103/PhysRevLett.43.103}{{\em Phys. Rev. Lett.}
  {\bfseries 43} (1979) 103}.

\bibitem{Shifman:1979if}
M.~A. Shifman, A.~I. Vainshtein, and V.~I. Zakharov, ``{Can Confinement Ensure
  Natural CP Invariance of Strong Interactions?},''
  \href{http://dx.doi.org/10.1016/0550-3213(80)90209-6}{{\em Nucl. Phys. B}
  {\bfseries 166} (1980) 493--506}.

\bibitem{Zhitnitsky:1980tq}
A.~R. Zhitnitsky, ``{On Possible Suppression of the Axion Hadron Interactions.
  (In Russian)},'' {\em Sov. J. Nucl. Phys.} {\bfseries 31} (1980) 260.

\bibitem{Dine:1981rt}
M.~Dine, W.~Fischler, and M.~Srednicki, ``{A Simple Solution to the Strong CP
  Problem with a Harmless Axion},''
  \href{http://dx.doi.org/10.1016/0370-2693(81)90590-6}{{\em Phys. Lett. B}
  {\bfseries 104} (1981) 199--202}.

\bibitem{Wilson:1973jj}
K.~G. Wilson and J.~B. Kogut, ``{The Renormalization group and the epsilon
  expansion},'' \href{http://dx.doi.org/10.1016/0370-1573(74)90023-4}{{\em
  Phys. Rept.} {\bfseries 12} (1974) 75--199}.

\bibitem{Giudice:2008bi}
G.~F. Giudice, ``{Naturally Speaking: The Naturalness Criterion and Physics at
  the LHC},'' \href{http://arxiv.org/abs/0801.2562}{{\ttfamily arXiv:0801.2562
  [hep-ph]}}.

\bibitem{Giudice:2013yca}
G.~F. Giudice, ``{Naturalness after LHC8},''
  \href{http://dx.doi.org/10.22323/1.180.0163}{{\em PoS} {\bfseries
  EPS-HEP2013} (2013) 163}, \href{http://arxiv.org/abs/1307.7879}{{\ttfamily
  arXiv:1307.7879 [hep-ph]}}.

\bibitem{Craig:2023lyt}
N.~Craig, ``{Naturalness: past, present, and future},''
  \href{http://dx.doi.org/10.1140/epjc/s10052-023-11928-7}{{\em Eur. Phys. J.
  C} {\bfseries 83} no.~9, (2023) 825},
  \href{http://arxiv.org/abs/2205.05708}{{\ttfamily arXiv:2205.05708
  [hep-ph]}}.

\bibitem{Vissani:1997ys}
F.~Vissani, ``{Do experiments suggest a hierarchy problem?},''
  \href{http://dx.doi.org/10.1103/PhysRevD.57.7027}{{\em Phys. Rev. D}
  {\bfseries 57} (1998) 7027--7030},
  \href{http://arxiv.org/abs/hep-ph/9709409}{{\ttfamily arXiv:hep-ph/9709409}}.

\bibitem{Shaposhnikov:2007nj}
M.~Shaposhnikov, ``{Is there a new physics between electroweak and Planck
  scales?},'' in {\em {Astroparticle Physics: Current Issues, 2007 (APCI07)}}.
\newblock 8, 2007.
\newblock \href{http://arxiv.org/abs/0708.3550}{{\ttfamily arXiv:0708.3550
  [hep-th]}}.

\bibitem{Farina:2013mla}
M.~Farina, D.~Pappadopulo, and A.~Strumia, ``{A modified naturalness principle
  and its experimental tests},''
  \href{http://dx.doi.org/10.1007/JHEP08(2013)022}{{\em JHEP} {\bfseries 08}
  (2013) 022}, \href{http://arxiv.org/abs/1303.7244}{{\ttfamily arXiv:1303.7244
  [hep-ph]}}.

\bibitem{tHooft:1972tcz}
G.~'t~Hooft and M.~J.~G. Veltman, ``{Regularization and Renormalization of
  Gauge Fields},'' \href{http://dx.doi.org/10.1016/0550-3213(72)90279-9}{{\em
  Nucl. Phys. B} {\bfseries 44} (1972) 189--213}.

\bibitem{Mooij:2021ojy}
S.~Mooij and M.~Shaposhnikov, ``{QFT without infinities and hierarchy
  problem},'' \href{http://dx.doi.org/10.1016/j.nuclphysb.2023.116172}{{\em
  Nucl. Phys. B} {\bfseries 990} (2023) 116172},
  \href{http://arxiv.org/abs/2110.05175}{{\ttfamily arXiv:2110.05175
  [hep-th]}}.

\bibitem{Mooij:2021lbc}
S.~Mooij and M.~Shaposhnikov, ``{Finite Callan-Symanzik renormalisation for
  multiple scalar fields},''
  \href{http://dx.doi.org/10.1016/j.nuclphysb.2023.116176}{{\em Nucl. Phys. B}
  {\bfseries 990} (2023) 116176},
  \href{http://arxiv.org/abs/2110.15925}{{\ttfamily arXiv:2110.15925
  [hep-th]}}.

\bibitem{Mooij:2024rys}
S.~Mooij and M.~Shaposhnikov, ``{Effective potential in finite formulation of
  QFT},'' \href{http://dx.doi.org/10.1016/j.nuclphysb.2024.116642}{{\em Nucl.
  Phys. B} {\bfseries 1006} (2024) 116642},
  \href{http://arxiv.org/abs/2408.00389}{{\ttfamily arXiv:2408.00389
  [hep-th]}}.

\bibitem{Ageeva:2024qie}
Y.~Ageeva, P.~Petrov, and M.~Shaposhnikov, ``{Nonrenormalizable theories and
  finite formulation of QFT},''
  \href{http://dx.doi.org/10.1103/PhysRevD.111.025020}{{\em Phys. Rev. D}
  {\bfseries 111} no.~2, (2025) 025020},
  \href{http://arxiv.org/abs/2409.15036}{{\ttfamily arXiv:2409.15036
  [hep-th]}}.

\bibitem{Coleman:1988cy}
S.~R. Coleman, ``{Black holes as red herrings: Topological fluctuations and the
  loss of quantum coherence},''
  \href{http://dx.doi.org/10.1016/0550-3213(88)90110-1}{{\em Nucl. Phys. B}
  {\bfseries 307} (1988) 867--882}.

\bibitem{Coleman:1988tj}
S.~R. Coleman, ``{Why There Is Nothing Rather Than Something: A Theory of the
  Cosmological Constant},''
  \href{http://dx.doi.org/10.1016/0550-3213(88)90097-1}{{\em Nucl. Phys. B}
  {\bfseries 310} (1988) 643--668}.

\bibitem{Hebecker:2018ofv}
A.~Hebecker, T.~Mikhail, and P.~Soler, ``{Euclidean wormholes, baby universes,
  and their impact on particle physics and cosmology},''
  \href{http://dx.doi.org/10.3389/fspas.2018.00035}{{\em Front. Astron. Space
  Sci.} {\bfseries 5} (2018) 35},
  \href{http://arxiv.org/abs/1807.00824}{{\ttfamily arXiv:1807.00824
  [hep-th]}}.

\bibitem{Kirzhnits:1972ut}
D.~A. Kirzhnits and A.~D. Linde, ``{Macroscopic Consequences of the Weinberg
  Model},'' \href{http://dx.doi.org/10.1016/0370-2693(72)90109-8}{{\em Phys.
  Lett. B} {\bfseries 42} (1972) 471--474}.

\bibitem{Weinberg:1974hy}
S.~Weinberg, ``{Gauge and Global Symmetries at High Temperature},''
  \href{http://dx.doi.org/10.1103/PhysRevD.9.3357}{{\em Phys. Rev. D}
  {\bfseries 9} (1974) 3357--3378}.

\bibitem{Shaposhnikov:1987tw}
M.~E. Shaposhnikov, ``{Baryon Asymmetry of the Universe in Standard Electroweak
  Theory},'' \href{http://dx.doi.org/10.1016/0550-3213(87)90127-1}{{\em Nucl.
  Phys. B} {\bfseries 287} (1987) 757--775}.

\bibitem{Shaposhnikov:1987pf}
M.~E. Shaposhnikov, ``{Structure of the High Temperature Gauge Ground State and
  Electroweak Production of the Baryon Asymmetry},''
  \href{http://dx.doi.org/10.1016/0550-3213(88)90373-2}{{\em Nucl. Phys. B}
  {\bfseries 299} (1988) 797--817}.

\bibitem{McLerran:1990zh}
L.~D. McLerran, M.~E. Shaposhnikov, N.~Turok, and M.~B. Voloshin, ``{Why the
  baryon asymmetry of the universe is approximately 10**-10},''
  \href{http://dx.doi.org/10.1016/0370-2693(91)91794-V}{{\em Phys. Lett. B}
  {\bfseries 256} (1991) 451--456}.

\bibitem{Cohen:1990it}
A.~G. Cohen, D.~B. Kaplan, and A.~E. Nelson, ``{Baryogenesis at the weak phase
  transition},'' \href{http://dx.doi.org/10.1016/0550-3213(91)90395-E}{{\em
  Nucl. Phys. B} {\bfseries 349} (1991) 727--742}.

\bibitem{Cohen:1990py}
A.~G. Cohen, D.~B. Kaplan, and A.~E. Nelson, ``{Weak scale baryogenesis},''
  \href{http://dx.doi.org/10.1016/0370-2693(90)90690-8}{{\em Phys. Lett. B}
  {\bfseries 245} (1990) 561--564}.

\bibitem{Turok:1990in}
N.~Turok and J.~Zadrozny, ``{Dynamical generation of baryons at the electroweak
  transition},'' \href{http://dx.doi.org/10.1103/PhysRevLett.65.2331}{{\em
  Phys. Rev. Lett.} {\bfseries 65} (1990) 2331--2334}.

\bibitem{Dine:1990fj}
M.~Dine, P.~Huet, R.~L. Singleton, Jr, and L.~Susskind, ``{Creating the baryon
  asymmetry at the electroweak phase transition},''
  \href{http://dx.doi.org/10.1016/0370-2693(91)91905-B}{{\em Phys. Lett. B}
  {\bfseries 257} (1991) 351--356}.

\bibitem{Nelson:1991ab}
A.~E. Nelson, D.~B. Kaplan, and A.~G. Cohen, ``{Why there is something rather
  than nothing: Matter from weak interactions},''
  \href{http://dx.doi.org/10.1016/0550-3213(92)90440-M}{{\em Nucl. Phys. B}
  {\bfseries 373} (1992) 453--478}.

\bibitem{vandeVis:2025efm}
J.~van~de Vis, J.~de~Vries, and M.~Postma, ``{Bubble Trouble: a Review on
  Electroweak Baryogenesis},''
  \href{http://arxiv.org/abs/2508.09989}{{\ttfamily arXiv:2508.09989
  [hep-ph]}}.

\bibitem{Vachaspati:2020blt}
T.~Vachaspati, ``{Progress on cosmological magnetic fields},''
  \href{http://dx.doi.org/10.1088/1361-6633/ac03a9}{{\em Rept. Prog. Phys.}
  {\bfseries 84} no.~7, (2021) 074901},
  \href{http://arxiv.org/abs/2010.10525}{{\ttfamily arXiv:2010.10525
  [astro-ph.CO]}}.

\bibitem{Caprini:2009yp}
C.~Caprini, R.~Durrer, and G.~Servant, ``{The stochastic gravitational wave
  background from turbulence and magnetic fields generated by a first-order
  phase transition},''
  \href{http://dx.doi.org/10.1088/1475-7516/2009/12/024}{{\em JCAP} {\bfseries
  12} (2009) 024}, \href{http://arxiv.org/abs/0909.0622}{{\ttfamily
  arXiv:0909.0622 [astro-ph.CO]}}.

\bibitem{Caprini:2019egz}
C.~Caprini {\em et~al.}, ``{Detecting gravitational waves from cosmological
  phase transitions with LISA: an update},''
  \href{http://dx.doi.org/10.1088/1475-7516/2020/03/024}{{\em JCAP} {\bfseries
  03} (2020) 024}, \href{http://arxiv.org/abs/1910.13125}{{\ttfamily
  arXiv:1910.13125 [astro-ph.CO]}}.

\bibitem{Caprini:2025trt}
C.~Caprini, ``{Probing Beyond the Standard Model with Gravitational Waves from
  Phase Transitions},''
  \href{http://dx.doi.org/10.53941/hihep.2025.100022}{{\em HiHEP} {\bfseries 1}
  no.~3, (2025) 22}, \href{http://arxiv.org/abs/2602.02861}{{\ttfamily
  arXiv:2602.02861 [astro-ph.CO]}}.

\bibitem{Elitzur:1975im}
S.~Elitzur, ``{Impossibility of Spontaneously Breaking Local Symmetries},''
  \href{http://dx.doi.org/10.1103/PhysRevD.12.3978}{{\em Phys. Rev. D}
  {\bfseries 12} (1975) 3978--3982}.

\bibitem{Banks:1979fi}
T.~Banks and E.~Rabinovici, ``{Finite Temperature Behavior of the Lattice
  Abelian Higgs Model},''
  \href{http://dx.doi.org/10.1016/0550-3213(79)90064-6}{{\em Nucl. Phys. B}
  {\bfseries 160} (1979) 349--379}.

\bibitem{Fradkin:1978dv}
E.~H. Fradkin and S.~H. Shenker, ``{Phase Diagrams of Lattice Gauge Theories
  with Higgs Fields},'' \href{http://dx.doi.org/10.1103/PhysRevD.19.3682}{{\em
  Phys. Rev. D} {\bfseries 19} (1979) 3682--3697}.

\bibitem{Farakos:1994kx}
K.~Farakos, K.~Kajantie, K.~Rummukainen, and M.~E. Shaposhnikov, ``{3-D physics
  and the electroweak phase transition: Perturbation theory},''
  \href{http://dx.doi.org/10.1016/0550-3213(94)90173-2}{{\em Nucl. Phys. B}
  {\bfseries 425} (1994) 67--109},
  \href{http://arxiv.org/abs/hep-ph/9404201}{{\ttfamily arXiv:hep-ph/9404201}}.

\bibitem{Kajantie:1996mn}
K.~Kajantie, M.~Laine, K.~Rummukainen, and M.~E. Shaposhnikov, ``{Is there a~
  hot electroweak phase transition at $m_H \gtrsim m_W$?},''
  \href{http://dx.doi.org/10.1103/PhysRevLett.77.2887}{{\em Phys. Rev. Lett.}
  {\bfseries 77} (1996) 2887--2890},
  \href{http://arxiv.org/abs/hep-ph/9605288}{{\ttfamily arXiv:hep-ph/9605288}}.

\bibitem{Kajantie:1995dw}
K.~Kajantie, M.~Laine, K.~Rummukainen, and M.~E. Shaposhnikov, ``{Generic rules
  for high temperature dimensional reduction and their application to the
  standard model},'' \href{http://dx.doi.org/10.1016/0550-3213(95)00549-8}{{\em
  Nucl. Phys. B} {\bfseries 458} (1996) 90--136},
  \href{http://arxiv.org/abs/hep-ph/9508379}{{\ttfamily arXiv:hep-ph/9508379}}.

\bibitem{Kajantie:1995kf}
K.~Kajantie, M.~Laine, K.~Rummukainen, and M.~E. Shaposhnikov, ``{The
  Electroweak phase transition: A Nonperturbative analysis},''
  \href{http://dx.doi.org/10.1016/0550-3213(96)00052-1}{{\em Nucl. Phys. B}
  {\bfseries 466} (1996) 189--258},
  \href{http://arxiv.org/abs/hep-lat/9510020}{{\ttfamily
  arXiv:hep-lat/9510020}}.

\bibitem{Fodor:1994dm}
Z.~Fodor, J.~Hein, K.~Jansen, A.~Jaster, I.~Montvay, and F.~Csikor,
  ``{Numerical simulations and the strength of the electroweak phase
  transition},'' \href{http://dx.doi.org/10.1016/0370-2693(94)90706-4}{{\em
  Phys. Lett. B} {\bfseries 334} (1994) 405--411},
  \href{http://arxiv.org/abs/hep-lat/9405021}{{\ttfamily
  arXiv:hep-lat/9405021}}.

\bibitem{Fodor:1994sj}
Z.~Fodor, J.~Hein, K.~Jansen, A.~Jaster, and I.~Montvay, ``{Simulating the
  electroweak phase transition in the SU(2) Higgs model},''
  \href{http://dx.doi.org/10.1016/0550-3213(95)00038-T}{{\em Nucl. Phys. B}
  {\bfseries 439} (1995) 147--186},
  \href{http://arxiv.org/abs/hep-lat/9409017}{{\ttfamily
  arXiv:hep-lat/9409017}}.

\bibitem{Laine:2012jy}
M.~Laine, G.~Nardini, and K.~Rummukainen, ``{Lattice study of an electroweak
  phase transition at $m_h \backsimeq$ 126 GeV},''
  \href{http://dx.doi.org/10.1088/1475-7516/2013/01/011}{{\em JCAP} {\bfseries
  01} (2013) 011}, \href{http://arxiv.org/abs/1211.7344}{{\ttfamily
  arXiv:1211.7344 [hep-ph]}}.

\bibitem{DOnofrio:2015gop}
M.~D'Onofrio and K.~Rummukainen, ``{Standard model cross-over on the
  lattice},'' \href{http://dx.doi.org/10.1103/PhysRevD.93.025003}{{\em Phys.
  Rev. D} {\bfseries 93} no.~2, (2016) 025003},
  \href{http://arxiv.org/abs/1508.07161}{{\ttfamily arXiv:1508.07161
  [hep-ph]}}.

\bibitem{Rummukainen:1998as}
K.~Rummukainen, M.~Tsypin, K.~Kajantie, M.~Laine, and M.~E. Shaposhnikov,
  ``{The Universality class of the electroweak theory},''
  \href{http://dx.doi.org/10.1016/S0550-3213(98)00494-5}{{\em Nucl. Phys. B}
  {\bfseries 532} (1998) 283--314},
  \href{http://arxiv.org/abs/hep-lat/9805013}{{\ttfamily
  arXiv:hep-lat/9805013}}.

\bibitem{Linde:1980ts}
A.~D. Linde, ``{Infrared Problem in Thermodynamics of the Yang-Mills Gas},''
  \href{http://dx.doi.org/10.1016/0370-2693(80)90769-8}{{\em Phys. Lett. B}
  {\bfseries 96} (1980) 289--292}.

\bibitem{Bruggisser:2018mrt}
S.~Bruggisser, B.~Von~Harling, O.~Matsedonskyi, and G.~Servant, ``{Electroweak
  Phase Transition and Baryogenesis in Composite Higgs Models},''
  \href{http://dx.doi.org/10.1007/JHEP12(2018)099}{{\em JHEP} {\bfseries 12}
  (2018) 099}, \href{http://arxiv.org/abs/1804.07314}{{\ttfamily
  arXiv:1804.07314 [hep-ph]}}.

\bibitem{Cohen:1993nk}
A.~G. Cohen, D.~B. Kaplan, and A.~E. Nelson, ``{Progress in electroweak
  baryogenesis},''
  \href{http://dx.doi.org/10.1146/annurev.ns.43.120193.000331}{{\em Ann. Rev.
  Nucl. Part. Sci.} {\bfseries 43} (1993) 27--70},
  \href{http://arxiv.org/abs/hep-ph/9302210}{{\ttfamily arXiv:hep-ph/9302210}}.

\bibitem{Rubakov:1996vz}
V.~A. Rubakov and M.~E. Shaposhnikov, ``{Electroweak baryon number
  nonconservation in the early universe and in high-energy collisions},''
  \href{http://dx.doi.org/10.1070/PU1996v039n05ABEH000145}{{\em Usp. Fiz. Nauk}
  {\bfseries 166} (1996) 493--537},
  \href{http://arxiv.org/abs/hep-ph/9603208}{{\ttfamily arXiv:hep-ph/9603208}}.

\bibitem{Trodden:1998ym}
M.~Trodden, ``{Electroweak baryogenesis},''
  \href{http://dx.doi.org/10.1103/RevModPhys.71.1463}{{\em Rev. Mod. Phys.}
  {\bfseries 71} (1999) 1463--1500},
  \href{http://arxiv.org/abs/hep-ph/9803479}{{\ttfamily arXiv:hep-ph/9803479}}.

\bibitem{Riotto:1999yt}
A.~Riotto and M.~Trodden, ``{Recent progress in baryogenesis},''
  \href{http://dx.doi.org/10.1146/annurev.nucl.49.1.35}{{\em Ann. Rev. Nucl.
  Part. Sci.} {\bfseries 49} (1999) 35--75},
  \href{http://arxiv.org/abs/hep-ph/9901362}{{\ttfamily arXiv:hep-ph/9901362}}.

\bibitem{Dine:2003ax}
M.~Dine and A.~Kusenko, ``{The Origin of the matter - antimatter asymmetry},''
  \href{http://dx.doi.org/10.1103/RevModPhys.76.1}{{\em Rev. Mod. Phys.}
  {\bfseries 76} (2003) 1},
  \href{http://arxiv.org/abs/hep-ph/0303065}{{\ttfamily arXiv:hep-ph/0303065}}.

\bibitem{Morrissey:2012db}
D.~E. Morrissey and M.~J. Ramsey-Musolf, ``{Electroweak baryogenesis},''
  \href{http://dx.doi.org/10.1088/1367-2630/14/12/125003}{{\em New J. Phys.}
  {\bfseries 14} (2012) 125003},
  \href{http://arxiv.org/abs/1206.2942}{{\ttfamily arXiv:1206.2942 [hep-ph]}}.

\bibitem{Garbrecht:2018mrp}
B.~Garbrecht, ``{Why is there more matter than antimatter? Calculational
  methods for leptogenesis and electroweak baryogenesis},''
  \href{http://dx.doi.org/10.1016/j.ppnp.2019.103727}{{\em Prog. Part. Nucl.
  Phys.} {\bfseries 110} (2020) 103727},
  \href{http://arxiv.org/abs/1812.02651}{{\ttfamily arXiv:1812.02651
  [hep-ph]}}.

\bibitem{Bruggisser:2022rdm}
S.~Bruggisser, B.~von Harling, O.~Matsedonskyi, and G.~Servant, ``{Status of
  electroweak baryogenesis in minimal composite Higgs},''
  \href{http://dx.doi.org/10.1007/JHEP08(2023)012}{{\em JHEP} {\bfseries 08}
  (2023) 012}, \href{http://arxiv.org/abs/2212.11953}{{\ttfamily
  arXiv:2212.11953 [hep-ph]}}.

\bibitem{Domcke:2025itc}
V.~Domcke, Y.~Ema, and T.~Patel, ``{Spontaneous magnetogenesis at the
  electroweak phase transition},''
  \href{http://dx.doi.org/10.1088/1475-7516/2025/09/023}{{\em JCAP} {\bfseries
  09} (2025) 023}, \href{http://arxiv.org/abs/2505.10637}{{\ttfamily
  arXiv:2505.10637 [hep-ph]}}.

\bibitem{Bhusal:2025lvm}
N.~Bhusal, S.~Blasi, M.~Cataldi, A.~Chatrchyan, M.~Gorghetto, and G.~Servant,
  ``{Standard Model Baryon Number Violation at Zero Temperature from Higgs
  Bubble Collisions},'' \href{http://arxiv.org/abs/2508.21825}{{\ttfamily
  arXiv:2508.21825 [hep-ph]}}.

\bibitem{vonHarling:2023dfl}
B.~von Harling, O.~Matsedonskyi, and G.~Servant, ``{High-temperature
  electroweak baryogenesis with composite Higgs},''
  \href{http://dx.doi.org/10.1007/JHEP12(2023)138}{{\em JHEP} {\bfseries 12}
  (2023) 138}, \href{http://arxiv.org/abs/2307.14426}{{\ttfamily
  arXiv:2307.14426 [hep-ph]}}.

\bibitem{Super-Kamiokande:1998kpq}
{\bfseries Super-Kamiokande} Collaboration, Y.~Fukuda {\em et~al.}, ``{Evidence
  for oscillation of atmospheric neutrinos},''
  \href{http://dx.doi.org/10.1103/PhysRevLett.81.1562}{{\em Phys. Rev. Lett.}
  {\bfseries 81} (1998) 1562--1567},
  \href{http://arxiv.org/abs/hep-ex/9807003}{{\ttfamily arXiv:hep-ex/9807003}}.

\bibitem{Kuzmin:1985mm}
V.~A. Kuzmin, V.~A. Rubakov, and M.~E. Shaposhnikov, ``{On the Anomalous
  Electroweak Baryon Number Nonconservation in the Early Universe},''
  \href{http://dx.doi.org/10.1016/0370-2693(85)91028-7}{{\em Phys. Lett. B}
  {\bfseries 155} (1985) 36}.

\bibitem{Adler:1969gk}
S.~L. Adler, ``{Axial vector vertex in spinor electrodynamics},''
  \href{http://dx.doi.org/10.1103/PhysRev.177.2426}{{\em Phys. Rev.} {\bfseries
  177} (1969) 2426--2438}.

\bibitem{Bell:1969ts}
J.~S. Bell and R.~Jackiw, ``{A PCAC puzzle: $\pi^0 \to \gamma \gamma$ in the
  $\sigma$ model},'' \href{http://dx.doi.org/10.1007/BF02823296}{{\em Nuovo
  Cim. A} {\bfseries 60} (1969) 47--61}.

\bibitem{tHooft:1976rip}
G.~'t~Hooft, ``{Symmetry Breaking Through Bell-Jackiw Anomalies},''
  \href{http://dx.doi.org/10.1103/PhysRevLett.37.8}{{\em Phys. Rev. Lett.}
  {\bfseries 37} (1976) 8--11}.

\bibitem{Sakharov:1967dj}
A.~D. Sakharov, ``{Violation of CP Invariance, C asymmetry, and baryon
  asymmetry of the universe},''
  \href{http://dx.doi.org/10.1070/PU1991v034n05ABEH002497}{{\em Pisma Zh. Eksp.
  Teor. Fiz.} {\bfseries 5} (1967) 32--35}.

\bibitem{tHooft:1976snw}
G.~'t~Hooft, ``{Computation of the Quantum Effects Due to a Four-Dimensional
  Pseudoparticle},'' \href{http://dx.doi.org/10.1103/PhysRevD.14.3432}{{\em
  Phys. Rev. D} {\bfseries 14} (1976) 3432--3450}. [Erratum: Phys.Rev.D 18,
  2199 (1978)].

\bibitem{Klinkhamer:1984di}
F.~R. Klinkhamer and N.~S. Manton, ``{A Saddle Point Solution in the
  Weinberg-Salam Theory},''
  \href{http://dx.doi.org/10.1103/PhysRevD.30.2212}{{\em Phys. Rev. D}
  {\bfseries 30} (1984) 2212}.

\bibitem{DOnofrio:2014rug}
M.~D'Onofrio, K.~Rummukainen, and A.~Tranberg, ``{Sphaleron Rate in the Minimal
  Standard Model},''
  \href{http://dx.doi.org/10.1103/PhysRevLett.113.141602}{{\em Phys. Rev.
  Lett.} {\bfseries 113} no.~14, (2014) 141602},
  \href{http://arxiv.org/abs/1404.3565}{{\ttfamily arXiv:1404.3565 [hep-ph]}}.

\bibitem{Shaposhnikov:1986jp}
M.~E. Shaposhnikov, ``{Possible Appearance of the Baryon Asymmetry of the
  Universe in an Electroweak Theory},'' {\em JETP Lett.} {\bfseries 44} (1986)
  465--468.

\bibitem{Kuzmin:1992up}
V.~A. Kuzmin, M.~E. Shaposhnikov, and I.~I. Tkachev, ``{Strong CP violation,
  electroweak baryogenesis, and axionic dark matter},''
  \href{http://dx.doi.org/10.1103/PhysRevD.45.466}{{\em Phys. Rev. D}
  {\bfseries 45} (1992) 466--475}.

\bibitem{Jarlskog:1985ht}
C.~Jarlskog, ``{Commutator of the Quark Mass Matrices in the Standard
  Electroweak Model and a Measure of Maximal CP Nonconservation},''
  \href{http://dx.doi.org/10.1103/PhysRevLett.55.1039}{{\em Phys. Rev. Lett.}
  {\bfseries 55} (1985) 1039}.

\bibitem{Nasser:1994bj}
S.~Nasser and N.~Turok, ``{Z condensation and standard model baryogenesis},''
  \href{http://arxiv.org/abs/hep-ph/9406270}{{\ttfamily arXiv:hep-ph/9406270}}.

\bibitem{Farrar:1993hn}
G.~R. Farrar and M.~E. Shaposhnikov, ``{Baryon asymmetry of the universe in the
  standard electroweak theory},''
  \href{http://dx.doi.org/10.1103/PhysRevD.50.774}{{\em Phys. Rev. D}
  {\bfseries 50} (1994) 774},
  \href{http://arxiv.org/abs/hep-ph/9305275}{{\ttfamily arXiv:hep-ph/9305275}}.

\bibitem{Kharzeev:2019rsy}
D.~Kharzeev, E.~Shuryak, and I.~Zahed, ``{Sphalerons, baryogenesis, and helical
  magnetogenesis in the electroweak transition of the minimal standard
  model},'' \href{http://dx.doi.org/10.1103/PhysRevD.102.073003}{{\em Phys.
  Rev. D} {\bfseries 102} no.~7, (2020) 073003},
  \href{http://arxiv.org/abs/1906.04080}{{\ttfamily arXiv:1906.04080
  [hep-ph]}}.

\bibitem{Ambjorn:1988gf}
J.~Ambjorn, M.~L. Laursen, and M.~E. Shaposhnikov, ``{Baryon Asymmetry
  Generation in the Electroweak Theory: A Lattice Study},''
  \href{http://dx.doi.org/10.1016/0550-3213(89)90043-6}{{\em Nucl. Phys. B}
  {\bfseries 316} (1989) 483--508}.

\bibitem{Gavela:1994dt}
M.~B. Gavela, P.~Hernandez, J.~Orloff, O.~Pene, and C.~Quimbay, ``{Standard
  model CP violation and baryon asymmetry. Part 2: Finite temperature},''
  \href{http://dx.doi.org/10.1016/0550-3213(94)00410-2}{{\em Nucl. Phys. B}
  {\bfseries 430} (1994) 382--426},
  \href{http://arxiv.org/abs/hep-ph/9406289}{{\ttfamily arXiv:hep-ph/9406289}}.

\bibitem{Huet:1994jb}
P.~Huet and E.~Sather, ``{Electroweak baryogenesis and standard model CP
  violation},'' \href{http://dx.doi.org/10.1103/PhysRevD.51.379}{{\em Phys.
  Rev. D} {\bfseries 51} (1995) 379--394},
  \href{http://arxiv.org/abs/hep-ph/9404302}{{\ttfamily arXiv:hep-ph/9404302}}.

\bibitem{Hong:2023zrf}
M.~Hong, K.~Kamada, and J.~Yokoyama, ``{Baryogenesis from sphaleron
  decoupling},'' \href{http://dx.doi.org/10.1103/PhysRevD.108.063502}{{\em
  Phys. Rev. D} {\bfseries 108} no.~6, (2023) 063502},
  \href{http://arxiv.org/abs/2304.13999}{{\ttfamily arXiv:2304.13999
  [hep-ph]}}.

\bibitem{Dolgov:1979mz}
A.~D. Dolgov, ``{Baryon asymmetry of the universe and violation of the
  thermodynamic equilibrium},'' {\em Pisma Zh. Eksp. Teor. Fiz.} {\bfseries 29}
  (1979) 254--258.

\bibitem{Dodelson:1993je}
S.~Dodelson and L.~M. Widrow, ``{Sterile-neutrinos as dark matter},''
  \href{http://dx.doi.org/10.1103/PhysRevLett.72.17}{{\em Phys. Rev. Lett.}
  {\bfseries 72} (1994) 17--20},
  \href{http://arxiv.org/abs/hep-ph/9303287}{{\ttfamily arXiv:hep-ph/9303287}}.

\bibitem{Boyarsky:2009ix}
A.~Boyarsky, O.~Ruchayskiy, and M.~Shaposhnikov, ``{The Role of sterile
  neutrinos in cosmology and astrophysics},''
  \href{http://dx.doi.org/10.1146/annurev.nucl.010909.083654}{{\em Ann. Rev.
  Nucl. Part. Sci.} {\bfseries 59} (2009) 191--214},
  \href{http://arxiv.org/abs/0901.0011}{{\ttfamily arXiv:0901.0011 [hep-ph]}}.

\bibitem{Kusenko:2009up}
A.~Kusenko, ``{Sterile neutrinos: The Dark side of the light fermions},''
  \href{http://dx.doi.org/10.1016/j.physrep.2009.07.004}{{\em Phys. Rept.}
  {\bfseries 481} (2009) 1--28},
  \href{http://arxiv.org/abs/0906.2968}{{\ttfamily arXiv:0906.2968 [hep-ph]}}.

\bibitem{Boyarsky:2018tvu}
A.~Boyarsky, M.~Drewes, T.~Lasserre, S.~Mertens, and O.~Ruchayskiy, ``{Sterile
  neutrino Dark Matter},''
  \href{http://dx.doi.org/10.1016/j.ppnp.2018.07.004}{{\em Prog. Part. Nucl.
  Phys.} {\bfseries 104} (2019) 1--45},
  \href{http://arxiv.org/abs/1807.07938}{{\ttfamily arXiv:1807.07938
  [hep-ph]}}.

\bibitem{Boyarsky:2006jm}
A.~Boyarsky, A.~Neronov, O.~Ruchayskiy, and M.~Shaposhnikov, ``{The Masses of
  active neutrinos in the nuMSM from X-ray astronomy},''
  \href{http://dx.doi.org/10.1134/S0021364006040011}{{\em JETP Lett.}
  {\bfseries 83} (2006) 133--135},
  \href{http://arxiv.org/abs/hep-ph/0601098}{{\ttfamily arXiv:hep-ph/0601098}}.

\bibitem{Fukugita:1986hr}
M.~Fukugita and T.~Yanagida, ``{Baryogenesis Without Grand Unification},''
  \href{http://dx.doi.org/10.1016/0370-2693(86)91126-3}{{\em Phys. Lett. B}
  {\bfseries 174} (1986) 45--47}.

\bibitem{Akhmedov:1998qx}
E.~K. Akhmedov, V.~A. Rubakov, and A.~Y. Smirnov, ``{Baryogenesis via neutrino
  oscillations},'' \href{http://dx.doi.org/10.1103/PhysRevLett.81.1359}{{\em
  Phys. Rev. Lett.} {\bfseries 81} (1998) 1359--1362},
  \href{http://arxiv.org/abs/hep-ph/9803255}{{\ttfamily arXiv:hep-ph/9803255}}.

\bibitem{Klaric:2020phc}
J.~Klari{\'c}, M.~Shaposhnikov, and I.~Timiryasov, ``{Uniting Low-Scale
  Leptogenesis Mechanisms},''
  \href{http://dx.doi.org/10.1103/PhysRevLett.127.111802}{{\em Phys. Rev.
  Lett.} {\bfseries 127} no.~11, (2021) 111802},
  \href{http://arxiv.org/abs/2008.13771}{{\ttfamily arXiv:2008.13771
  [hep-ph]}}.

\bibitem{Abada:2018oly}
A.~Abada, G.~Arcadi, V.~Domcke, M.~Drewes, J.~Klaric, and M.~Lucente,
  ``{Low-scale leptogenesis with three heavy neutrinos},''
  \href{http://dx.doi.org/10.1007/JHEP01(2019)164}{{\em JHEP} {\bfseries 01}
  (2019) 164}, \href{http://arxiv.org/abs/1810.12463}{{\ttfamily
  arXiv:1810.12463 [hep-ph]}}.

\bibitem{Drewes:2024bla}
M.~Drewes, Y.~Georis, J.~Klari{\'c}, and A.~Wendels, ``{On the
  collider-testability of the type-I seesaw model with 3 right-handed
  neutrinos},'' \href{http://dx.doi.org/10.1007/JHEP03(2025)176}{{\em JHEP}
  {\bfseries 03} (2025) 176}, \href{http://arxiv.org/abs/2407.13620}{{\ttfamily
  arXiv:2407.13620 [hep-ph]}}.

\bibitem{Shaposhnikov:2006nn}
M.~Shaposhnikov, ``{A Possible symmetry of the nuMSM},''
  \href{http://dx.doi.org/10.1016/j.nuclphysb.2006.11.003}{{\em Nucl. Phys. B}
  {\bfseries 763} (2007) 49--59},
  \href{http://arxiv.org/abs/hep-ph/0605047}{{\ttfamily arXiv:hep-ph/0605047}}.

\bibitem{Alekhin:2015byh}
S.~Alekhin {\em et~al.}, ``{A facility to Search for Hidden Particles at the
  CERN SPS: the SHiP physics case},''
  \href{http://dx.doi.org/10.1088/0034-4885/79/12/124201}{{\em Rept. Prog.
  Phys.} {\bfseries 79} no.~12, (2016) 124201},
  \href{http://arxiv.org/abs/1504.04855}{{\ttfamily arXiv:1504.04855
  [hep-ph]}}.

\bibitem{FCC:2025lpp}
{\bfseries FCC} Collaboration, M.~Benedikt {\em et~al.}, ``{Future Circular
  Collider Feasibility Study Report: Volume 1, Physics, Experiments,
  Detectors},'' \href{http://dx.doi.org/10.1140/epjc/s10052-025-15077-x}{{\em
  Eur. Phys. J. C} {\bfseries 85} no.~12, (2025) 1468},
  \href{http://arxiv.org/abs/2505.00272}{{\ttfamily arXiv:2505.00272
  [hep-ex]}}.

\bibitem{Gorbunov:2007ak}
D.~Gorbunov and M.~Shaposhnikov, ``{How to find neutral leptons of the
  $\nu$MSM?},'' \href{http://dx.doi.org/10.1088/1126-6708/2007/10/015}{{\em
  JHEP} {\bfseries 10} (2007) 015},
  \href{http://arxiv.org/abs/0705.1729}{{\ttfamily arXiv:0705.1729 [hep-ph]}}.
  [Erratum: JHEP 11, 101 (2013)].

\bibitem{Blondel:2014bra}
A.~Blondel, E.~Graverini, N.~Serra, and M.~Shaposhnikov, ``{Search for Heavy
  Right Handed Neutrinos at the FCC-ee},''
  \href{http://dx.doi.org/10.1016/j.nuclphysbps.2015.09.304}{{\em Nucl. Part.
  Phys. Proc.} {\bfseries 273-275} (2016) 1883--1890},
  \href{http://arxiv.org/abs/1411.5230}{{\ttfamily arXiv:1411.5230 [hep-ex]}}.

\bibitem{Hernandez:2016kel}
P.~Hern{\'a}ndez, M.~Kekic, J.~L{\'o}pez-Pav{\'o}n, J.~Racker, and J.~Salvado,
  ``{Testable Baryogenesis in Seesaw Models},''
  \href{http://dx.doi.org/10.1007/JHEP08(2016)157}{{\em JHEP} {\bfseries 08}
  (2016) 157}, \href{http://arxiv.org/abs/1606.06719}{{\ttfamily
  arXiv:1606.06719 [hep-ph]}}.

\bibitem{Drewes:2021nqr}
M.~Drewes, Y.~Georis, and J.~Klari{\'c}, ``{Mapping the Viable Parameter Space
  for Testable Leptogenesis},''
  \href{http://dx.doi.org/10.1103/PhysRevLett.128.051801}{{\em Phys. Rev.
  Lett.} {\bfseries 128} no.~5, (2022) 051801},
  \href{http://arxiv.org/abs/2106.16226}{{\ttfamily arXiv:2106.16226
  [hep-ph]}}.

\bibitem{Hernandez:2022ivz}
P.~Hernandez, J.~Lopez-Pavon, N.~Rius, and S.~Sandner, ``{Bounds on
  right-handed neutrino parameters from observable leptogenesis},''
  \href{http://dx.doi.org/10.1007/JHEP12(2022)012}{{\em JHEP} {\bfseries 12}
  (2022) 012}, \href{http://arxiv.org/abs/2207.01651}{{\ttfamily
  arXiv:2207.01651 [hep-ph]}}.

\bibitem{Sandner:2023tcg}
S.~Sandner, P.~Hernandez, J.~Lopez-Pavon, and N.~Rius, ``{Predicting the baryon
  asymmetry with degenerate right-handed neutrinos},''
  \href{http://dx.doi.org/10.1007/JHEP11(2023)153}{{\em JHEP} {\bfseries 11}
  (2023) 153}, \href{http://arxiv.org/abs/2305.14427}{{\ttfamily
  arXiv:2305.14427 [hep-ph]}}.

\bibitem{Anastopoulos:2025jyh}
C.~Anastopoulos {\em et~al.}, ``{LEP3: a high-luminosity $e^+e^-$ Higgs and
  electroweak factory in the LHC tunnel},''
  \href{http://dx.doi.org/10.1088/1361-6471/ae5728}{{\em J. Phys. G} {\bfseries
  53} no.~4, (2026) 040501}, \href{http://arxiv.org/abs/2504.00541}{{\ttfamily
  arXiv:2504.00541 [physics.acc-ph]}}.

\bibitem{Drewes:2025ohy}
M.~Drewes, E.~Shaposhnikova, and M.~Shaposhnikov, ``{A Possible Future Use of
  the LHC Tunnel},''
\newblock 3, 2025.
\newblock \href{http://arxiv.org/abs/2503.17081}{{\ttfamily arXiv:2503.17081
  [hep-ex]}}.

\bibitem{Bertone:2004pz}
G.~Bertone, D.~Hooper, and J.~Silk, ``{Particle dark matter: Evidence,
  candidates and constraints},''
  \href{http://dx.doi.org/10.1016/j.physrep.2004.08.031}{{\em Phys. Rept.}
  {\bfseries 405} (2005) 279--390},
  \href{http://arxiv.org/abs/hep-ph/0404175}{{\ttfamily arXiv:hep-ph/0404175}}.

\bibitem{Cirelli:2024ssz}
M.~Cirelli, A.~Strumia, and J.~Zupan, ``{Dark Matter},''
  \href{http://dx.doi.org/10.21468/SciPostPhysRev.1}{{\em SciPost Phys. Rev.}
  {\bfseries 1} (2026) }, \href{http://arxiv.org/abs/2406.01705}{{\ttfamily
  arXiv:2406.01705 [hep-ph]}}.

\bibitem{Shi:1998km}
X.-D. Shi and G.~M. Fuller, ``{A New dark matter candidate: Nonthermal sterile
  neutrinos},'' \href{http://dx.doi.org/10.1103/PhysRevLett.82.2832}{{\em Phys.
  Rev. Lett.} {\bfseries 82} (1999) 2832--2835},
  \href{http://arxiv.org/abs/astro-ph/9810076}{{\ttfamily
  arXiv:astro-ph/9810076}}.

\bibitem{Abazajian:2001nj}
K.~Abazajian, G.~M. Fuller, and M.~Patel, ``{Sterile neutrino hot, warm, and
  cold dark matter},'' \href{http://dx.doi.org/10.1103/PhysRevD.64.023501}{{\em
  Phys. Rev. D} {\bfseries 64} (2001) 023501},
  \href{http://arxiv.org/abs/astro-ph/0101524}{{\ttfamily
  arXiv:astro-ph/0101524}}.

\bibitem{Abazajian:2001vt}
K.~Abazajian, G.~M. Fuller, and W.~H. Tucker, ``{Direct detection of warm dark
  matter in the X-ray},'' \href{http://dx.doi.org/10.1086/323867}{{\em
  Astrophys. J.} {\bfseries 562} (2001) 593--604},
  \href{http://arxiv.org/abs/astro-ph/0106002}{{\ttfamily
  arXiv:astro-ph/0106002}}.

\bibitem{Abdullahi:2022jlv}
A.~M. Abdullahi {\em et~al.}, ``{The present and future status of heavy neutral
  leptons},'' \href{http://dx.doi.org/10.1088/1361-6471/ac98f9}{{\em J. Phys.
  G} {\bfseries 50} no.~2, (2023) 020501},
  \href{http://arxiv.org/abs/2203.08039}{{\ttfamily arXiv:2203.08039
  [hep-ph]}}.

\bibitem{Gorbunov:2025nqs}
D.~Gorbunov, D.~Kalashnikov, and G.~Krugan, ``{Playing with lepton asymmetry at
  the resonant production of sterile neutrino dark matter},''
  \href{http://dx.doi.org/10.1016/j.physletb.2025.139750}{{\em Phys. Lett. B}
  {\bfseries 868} (2025) 139750},
  \href{http://arxiv.org/abs/2502.17374}{{\ttfamily arXiv:2502.17374
  [hep-ph]}}.

\bibitem{Akita:2025txo}
K.~Akita, K.~Hamaguchi, and M.~Ovchynnikov, ``{Maximal Parameter Space of
  Sterile Neutrino Dark Matter with Lepton Asymmetries},''
  \href{http://dx.doi.org/10.1103/d2qg-clrz}{{\em Phys. Rev. Lett.} {\bfseries
  136} no.~11, (2026) 111001},
  \href{http://arxiv.org/abs/2507.20659}{{\ttfamily arXiv:2507.20659
  [hep-ph]}}.

\bibitem{Vogel:2025aut}
C.~M. Vogel, H.~Garc{\'\i}a~Escudero, J.~Froustey, and K.~N. Abazajian,
  ``{Return of the lepton number: Sterile neutrino dark matter production and
  the revival of the Shi-Fuller mechanism},''
  \href{http://dx.doi.org/10.1103/nys6-tkhl}{{\em Phys. Rev. D} {\bfseries 112}
  no.~12, (2025) 123508}, \href{http://arxiv.org/abs/2507.18752}{{\ttfamily
  arXiv:2507.18752 [hep-ph]}}.

\bibitem{Shaposhnikov:2023hrx}
M.~Shaposhnikov and A.~Y. Smirnov, ``{Sterile neutrino dark matter,
  matter-antimatter separation, and the QCD phase transition},''
  \href{http://dx.doi.org/10.1103/PhysRevD.110.063520}{{\em Phys. Rev. D}
  {\bfseries 110} no.~6, (2024) 063520},
  \href{http://arxiv.org/abs/2309.13376}{{\ttfamily arXiv:2309.13376
  [hep-ph]}}.

\bibitem{Shaposhnikov:2008pf}
M.~Shaposhnikov, ``{The nuMSM, leptonic asymmetries, and properties of singlet
  fermions},'' \href{http://dx.doi.org/10.1088/1126-6708/2008/08/008}{{\em
  JHEP} {\bfseries 08} (2008) 008},
  \href{http://arxiv.org/abs/0804.4542}{{\ttfamily arXiv:0804.4542 [hep-ph]}}.

\bibitem{Canetti:2012vf}
L.~Canetti, M.~Drewes, and M.~Shaposhnikov, ``{Sterile Neutrinos as the Origin
  of Dark and Baryonic Matter},''
  \href{http://dx.doi.org/10.1103/PhysRevLett.110.061801}{{\em Phys. Rev.
  Lett.} {\bfseries 110} no.~6, (2013) 061801},
  \href{http://arxiv.org/abs/1204.3902}{{\ttfamily arXiv:1204.3902 [hep-ph]}}.

\bibitem{Canetti:2012kh}
L.~Canetti, M.~Drewes, T.~Frossard, and M.~Shaposhnikov, ``{Dark Matter,
  Baryogenesis and Neutrino Oscillations from Right Handed Neutrinos},''
  \href{http://dx.doi.org/10.1103/PhysRevD.87.093006}{{\em Phys. Rev. D}
  {\bfseries 87} (2013) 093006},
  \href{http://arxiv.org/abs/1208.4607}{{\ttfamily arXiv:1208.4607 [hep-ph]}}.

\bibitem{Ghiglieri:2020ulj}
J.~Ghiglieri and M.~Laine, ``{Sterile neutrino dark matter via coinciding
  resonances},'' \href{http://dx.doi.org/10.1088/1475-7516/2020/07/012}{{\em
  JCAP} {\bfseries 07} (2020) 012},
  \href{http://arxiv.org/abs/2004.10766}{{\ttfamily arXiv:2004.10766
  [hep-ph]}}.

\bibitem{Shaposhnikov:2020aen}
M.~Shaposhnikov, A.~Shkerin, I.~Timiryasov, and S.~Zell, ``{Einstein-Cartan
  Portal to Dark Matter},''
  \href{http://dx.doi.org/10.1103/PhysRevLett.127.169901}{{\em Phys. Rev.
  Lett.} {\bfseries 126} no.~16, (2021) 161301},
  \href{http://arxiv.org/abs/2008.11686}{{\ttfamily arXiv:2008.11686
  [hep-ph]}}. [Erratum: Phys.Rev.Lett. 127, 169901 (2021)].

\bibitem{Tremaine:1979we}
S.~Tremaine and J.~E. Gunn, ``{Dynamical Role of Light Neutral Leptons in
  Cosmology},'' \href{http://dx.doi.org/10.1103/PhysRevLett.42.407}{{\em Phys.
  Rev. Lett.} {\bfseries 42} (1979) 407--410}.

\bibitem{Bezrukov:2025ttd}
F.~Bezrukov, D.~Gorbunov, and E.~Koreshkova, ``{Refining lower bounds on
  sterile neutrino dark matter mass from estimates of phase space densities in
  dwarf galaxies},'' \href{http://dx.doi.org/10.1142/s0217751x25400044}{{\em
  Int. J. Mod. Phys. A} {\bfseries 40} no.~33, (2025) 2540004},
  \href{http://arxiv.org/abs/2412.20585}{{\ttfamily arXiv:2412.20585
  [hep-ph]}}.

\bibitem{Viel:2005qj}
M.~Viel, J.~Lesgourgues, M.~G. Haehnelt, S.~Matarrese, and A.~Riotto,
  ``{Constraining warm dark matter candidates including sterile neutrinos and
  light gravitinos with WMAP and the Lyman-alpha forest},''
  \href{http://dx.doi.org/10.1103/PhysRevD.71.063534}{{\em Phys. Rev. D}
  {\bfseries 71} (2005) 063534},
  \href{http://arxiv.org/abs/astro-ph/0501562}{{\ttfamily
  arXiv:astro-ph/0501562}}.

\bibitem{Boyarsky:2008xj}
A.~Boyarsky, J.~Lesgourgues, O.~Ruchayskiy, and M.~Viel, ``{Lyman-alpha
  constraints on warm and on warm-plus-cold dark matter models},''
  \href{http://dx.doi.org/10.1088/1475-7516/2009/05/012}{{\em JCAP} {\bfseries
  05} (2009) 012}, \href{http://arxiv.org/abs/0812.0010}{{\ttfamily
  arXiv:0812.0010 [astro-ph]}}.

\bibitem{Viel:2013fqw}
M.~Viel, G.~D. Becker, J.~S. Bolton, and M.~G. Haehnelt, ``{Warm dark matter as
  a solution to the small scale crisis: New constraints from high redshift
  Lyman-{\ensuremath{\alpha}} forest data},''
  \href{http://dx.doi.org/10.1103/PhysRevD.88.043502}{{\em Phys. Rev. D}
  {\bfseries 88} (2013) 043502},
  \href{http://arxiv.org/abs/1306.2314}{{\ttfamily arXiv:1306.2314
  [astro-ph.CO]}}.

\bibitem{DES:2020fxi}
{\bfseries DES} Collaboration, E.~O. Nadler {\em et~al.}, ``{Milky Way
  Satellite Census. III. Constraints on Dark Matter Properties from
  Observations of Milky Way Satellite Galaxies},''
  \href{http://dx.doi.org/10.1103/PhysRevLett.126.091101}{{\em Phys. Rev.
  Lett.} {\bfseries 126} (2021) 091101},
  \href{http://arxiv.org/abs/2008.00022}{{\ttfamily arXiv:2008.00022
  [astro-ph.CO]}}.

\bibitem{Enzi:2020ieg}
W.~Enzi {\em et~al.}, ``{Joint constraints on thermal relic dark matter from
  strong gravitational lensing, the Ly{\,}{\ensuremath{\alpha}} forest, and
  Milky Way satellites},'' \href{http://dx.doi.org/10.1093/mnras/stab1960}{{\em
  Mon. Not. Roy. Astron. Soc.} {\bfseries 506} no.~4, (2021) 5848--5862},
  \href{http://arxiv.org/abs/2010.13802}{{\ttfamily arXiv:2010.13802
  [astro-ph.CO]}}.

\bibitem{Newton:2024jsy}
O.~Newton, M.~R. Lovell, C.~S. Frenk, A.~Jenkins, J.~C. Helly, S.~Cole, A.~J.
  Benson, and W.~A. Hellwing, ``{Constraints on the properties of
  {\ensuremath{\nu}}MSM dark matter using the satellite galaxies of the Milky
  Way},'' \href{http://dx.doi.org/10.1093/mnras/staf1223}{{\em Mon. Not. Roy.
  Astron. Soc.} {\bfseries 541} no.~4, (2025) 3713--3727},
  \href{http://arxiv.org/abs/2408.16042}{{\ttfamily arXiv:2408.16042
  [astro-ph.CO]}}.

\bibitem{Garzilli:2019qki}
A.~Garzilli, A.~Magalich, O.~Ruchayskiy, and A.~Boyarsky, ``{How to constrain
  warm dark matter with the Lyman-$\alpha$ forest},''
  \href{http://dx.doi.org/10.1093/mnras/stab192}{{\em Mon. Not. Roy. Astron.
  Soc.} {\bfseries 502} no.~2, (2021) 2356--2363},
  \href{http://arxiv.org/abs/1912.09397}{{\ttfamily arXiv:1912.09397
  [astro-ph.CO]}}.

\bibitem{Garzilli:2018jqh}
A.~Garzilli, A.~Magalich, T.~Theuns, C.~S. Frenk, C.~Weniger, O.~Ruchayskiy,
  and A.~Boyarsky, ``{The Lyman-$\alpha$ forest as a diagnostic of the nature
  of the dark matter},'' \href{http://dx.doi.org/10.1093/mnras/stz2188}{{\em
  Mon. Not. Roy. Astron. Soc.} {\bfseries 489} no.~3, (2019) 3456--3471},
  \href{http://arxiv.org/abs/1809.06585}{{\ttfamily arXiv:1809.06585
  [astro-ph.CO]}}.

\bibitem{Starobinsky:1980te}
A.~A. Starobinsky, ``{A New Type of Isotropic Cosmological Models Without
  Singularity},'' \href{http://dx.doi.org/10.1016/0370-2693(80)90670-X}{{\em
  Phys. Lett. B} {\bfseries 91} (1980) 99--102}.

\bibitem{Guth:1980zm}
A.~H. Guth, ``{The Inflationary Universe: A Possible Solution to the Horizon
  and Flatness Problems},''
  \href{http://dx.doi.org/10.1103/PhysRevD.23.347}{{\em Phys. Rev. D}
  {\bfseries 23} (1981) 347--356}.

\bibitem{Linde:1981mu}
A.~D. Linde, ``{A New Inflationary Universe Scenario: A Possible Solution of
  the Horizon, Flatness, Homogeneity, Isotropy and Primordial Monopole
  Problems},'' \href{http://dx.doi.org/10.1016/0370-2693(82)91219-9}{{\em Phys.
  Lett. B} {\bfseries 108} (1982) 389--393}.

\bibitem{Albrecht:1982wi}
A.~Albrecht and P.~J. Steinhardt, ``{Cosmology for Grand Unified Theories with
  Radiatively Induced Symmetry Breaking},''
  \href{http://dx.doi.org/10.1103/PhysRevLett.48.1220}{{\em Phys. Rev. Lett.}
  {\bfseries 48} (1982) 1220--1223}.

\bibitem{Mukhanov:1981xt}
V.~F. Mukhanov and G.~V. Chibisov, ``{Quantum Fluctuations and a Nonsingular
  Universe},'' {\em JETP Lett.} {\bfseries 33} (1981) 532--535.

\bibitem{Linde:1983gd}
A.~D. Linde, ``{Chaotic Inflation},''
  \href{http://dx.doi.org/10.1016/0370-2693(83)90837-7}{{\em Phys. Lett. B}
  {\bfseries 129} (1983) 177--181}.

\bibitem{Freese:1990rb}
K.~Freese, J.~A. Frieman, and A.~V. Olinto, ``{Natural inflation with pseudo
  Nambu-Goldstone bosons},''
  \href{http://dx.doi.org/10.1103/PhysRevLett.65.3233}{{\em Phys. Rev. Lett.}
  {\bfseries 65} (1990) 3233--3236}.

\bibitem{Pajer:2013fsa}
E.~Pajer and M.~Peloso, ``{A review of Axion Inflation in the era of Planck},''
  \href{http://dx.doi.org/10.1088/0264-9381/30/21/214002}{{\em Class. Quant.
  Grav.} {\bfseries 30} (2013) 214002},
  \href{http://arxiv.org/abs/1305.3557}{{\ttfamily arXiv:1305.3557 [hep-th]}}.

\bibitem{Kallosh:2013yoa}
R.~Kallosh, A.~Linde, and D.~Roest, ``{Superconformal Inflationary
  $\alpha$-Attractors},'' \href{http://dx.doi.org/10.1007/JHEP11(2013)198}{{\em
  JHEP} {\bfseries 11} (2013) 198},
  \href{http://arxiv.org/abs/1311.0472}{{\ttfamily arXiv:1311.0472 [hep-th]}}.

\bibitem{Martin:2013tda}
J.~Martin, C.~Ringeval, and V.~Vennin, ``{Encyclop\ae{}dia Inflationaris},''
  \href{http://dx.doi.org/10.1016/j.dark.2014.01.003}{{\em Phys. Dark Univ.}
  {\bfseries 5-6} (2014) 75--235},
  \href{http://arxiv.org/abs/1303.3787}{{\ttfamily arXiv:1303.3787
  [astro-ph.CO]}}.

\bibitem{Bezrukov:2007ep}
F.~L. Bezrukov and M.~Shaposhnikov, ``{The Standard Model Higgs boson as the
  inflaton},'' \href{http://dx.doi.org/10.1016/j.physletb.2007.11.072}{{\em
  Phys. Lett. B} {\bfseries 659} (2008) 703--706},
  \href{http://arxiv.org/abs/0710.3755}{{\ttfamily arXiv:0710.3755 [hep-th]}}.

\bibitem{Sakharov:1967pk}
A.~D. Sakharov, ``{Vacuum quantum fluctuations in curved space and the theory
  of gravitation},''
  \href{http://dx.doi.org/10.1070/PU1991v034n05ABEH002498}{{\em Dokl. Akad.
  Nauk Ser. Fiz.} {\bfseries 177} (1967) 70--71}.

\bibitem{Stelle:1976gc}
K.~S. Stelle, ``{Renormalization of Higher Derivative Quantum Gravity},''
  \href{http://dx.doi.org/10.1103/PhysRevD.16.953}{{\em Phys. Rev. D}
  {\bfseries 16} (1977) 953--969}.

\bibitem{Holdom:2024onr}
B.~Holdom, ``{Making sense of ghosts},''
  \href{http://dx.doi.org/10.1016/j.nuclphysb.2024.116696}{{\em Nucl. Phys. B}
  {\bfseries 1008} (2024) 116696},
  \href{http://arxiv.org/abs/2408.04089}{{\ttfamily arXiv:2408.04089
  [hep-th]}}.

\bibitem{Salvio:2015gsi}
A.~Salvio and A.~Strumia, ``{Quantum mechanics of 4-derivative theories},''
  \href{http://dx.doi.org/10.1140/epjc/s10052-016-4079-8}{{\em Eur. Phys. J. C}
  {\bfseries 76} no.~4, (2016) 227},
  \href{http://arxiv.org/abs/1512.01237}{{\ttfamily arXiv:1512.01237
  [hep-th]}}.

\bibitem{Spokoiny:1984bd}
B.~L. Spokoiny, ``{Inflation and Generation of Perturbations in Broken
  Symmetric Theory of Gravity},''
  \href{http://dx.doi.org/10.1016/0370-2693(84)90587-2}{{\em Phys. Lett. B}
  {\bfseries 147} (1984) 39--43}.

\bibitem{Salopek:1988qh}
D.~S. Salopek, J.~R. Bond, and J.~M. Bardeen, ``{Designing Density Fluctuation
  Spectra in Inflation},''
  \href{http://dx.doi.org/10.1103/PhysRevD.40.1753}{{\em Phys. Rev. D}
  {\bfseries 40} (1989) 1753}.

\bibitem{Kaiser:1994vs}
D.~I. Kaiser, ``{Primordial spectral indices from generalized Einstein
  theories},'' \href{http://dx.doi.org/10.1103/PhysRevD.52.4295}{{\em Phys.
  Rev. D} {\bfseries 52} (1995) 4295--4306},
  \href{http://arxiv.org/abs/astro-ph/9408044}{{\ttfamily
  arXiv:astro-ph/9408044}}.

\bibitem{Bezrukov:2008ut}
F.~Bezrukov, D.~Gorbunov, and M.~Shaposhnikov, ``{On initial conditions for the
  Hot Big Bang},'' \href{http://dx.doi.org/10.1088/1475-7516/2009/06/029}{{\em
  JCAP} {\bfseries 06} (2009) 029},
  \href{http://arxiv.org/abs/0812.3622}{{\ttfamily arXiv:0812.3622 [hep-ph]}}.

\bibitem{Garcia-Bellido:2008ycs}
J.~Garcia-Bellido, D.~G. Figueroa, and J.~Rubio, ``{Preheating in the Standard
  Model with the Higgs-Inflaton coupled to gravity},''
  \href{http://dx.doi.org/10.1103/PhysRevD.79.063531}{{\em Phys. Rev. D}
  {\bfseries 79} (2009) 063531},
  \href{http://arxiv.org/abs/0812.4624}{{\ttfamily arXiv:0812.4624 [hep-ph]}}.

\bibitem{DeCross:2015uza}
M.~P. DeCross, D.~I. Kaiser, A.~Prabhu, C.~Prescod-Weinstein, and E.~I.
  Sfakianakis, ``{Preheating after Multifield Inflation with Nonminimal
  Couplings, I: Covariant Formalism and Attractor Behavior},''
  \href{http://dx.doi.org/10.1103/PhysRevD.97.023526}{{\em Phys. Rev. D}
  {\bfseries 97} no.~2, (2018) 023526},
  \href{http://arxiv.org/abs/1510.08553}{{\ttfamily arXiv:1510.08553
  [astro-ph.CO]}}.

\bibitem{DeCross:2016fdz}
M.~P. DeCross, D.~I. Kaiser, A.~Prabhu, C.~Prescod-Weinstein, and E.~I.
  Sfakianakis, ``{Preheating after multifield inflation with nonminimal
  couplings, II: Resonance Structure},''
  \href{http://dx.doi.org/10.1103/PhysRevD.97.023527}{{\em Phys. Rev. D}
  {\bfseries 97} no.~2, (2018) 023527},
  \href{http://arxiv.org/abs/1610.08868}{{\ttfamily arXiv:1610.08868
  [astro-ph.CO]}}.

\bibitem{DeCross:2016cbs}
M.~P. DeCross, D.~I. Kaiser, A.~Prabhu, C.~Prescod-Weinstein, and E.~I.
  Sfakianakis, ``{Preheating after multifield inflation with nonminimal
  couplings, III: Dynamical spacetime results},''
  \href{http://dx.doi.org/10.1103/PhysRevD.97.023528}{{\em Phys. Rev. D}
  {\bfseries 97} no.~2, (2018) 023528},
  \href{http://arxiv.org/abs/1610.08916}{{\ttfamily arXiv:1610.08916
  [astro-ph.CO]}}.

\bibitem{Ema:2016dny}
Y.~Ema, R.~Jinno, K.~Mukaida, and K.~Nakayama, ``{Violent Preheating in
  Inflation with Nonminimal Coupling},''
  \href{http://dx.doi.org/10.1088/1475-7516/2017/02/045}{{\em JCAP} {\bfseries
  02} (2017) 045}, \href{http://arxiv.org/abs/1609.05209}{{\ttfamily
  arXiv:1609.05209 [hep-ph]}}.

\bibitem{Sfakianakis:2018lzf}
E.~I. Sfakianakis and J.~van~de Vis, ``{Preheating after Higgs Inflation:
  Self-Resonance and Gauge boson production},''
  \href{http://dx.doi.org/10.1103/PhysRevD.99.083519}{{\em Phys. Rev. D}
  {\bfseries 99} no.~8, (2019) 083519},
  \href{http://arxiv.org/abs/1810.01304}{{\ttfamily arXiv:1810.01304
  [hep-ph]}}.

\bibitem{AtacamaCosmologyTelescope:2025nti}
{\bfseries Atacama Cosmology Telescope} Collaboration, E.~Calabrese {\em
  et~al.}, ``{The Atacama Cosmology Telescope: DR6 constraints on extended
  cosmological models},''
  \href{http://dx.doi.org/10.1088/1475-7516/2025/11/063}{{\em JCAP} {\bfseries
  11} (2025) 063}, \href{http://arxiv.org/abs/2503.14454}{{\ttfamily
  arXiv:2503.14454 [astro-ph.CO]}}.

\bibitem{AtacamaCosmologyTelescope:2025blo}
{\bfseries Atacama Cosmology Telescope} Collaboration, T.~Louis {\em et~al.},
  ``{The Atacama Cosmology Telescope: DR6 power spectra, likelihoods and
  {\ensuremath{\Lambda}}CDM parameters},''
  \href{http://dx.doi.org/10.1088/1475-7516/2025/11/062}{{\em JCAP} {\bfseries
  11} (2025) 062}, \href{http://arxiv.org/abs/2503.14452}{{\ttfamily
  arXiv:2503.14452 [astro-ph.CO]}}.

\bibitem{DESI:2024uvr}
{\bfseries DESI} Collaboration, A.~G. Adame {\em et~al.}, ``{DESI 2024 III:
  baryon acoustic oscillations from galaxies and quasars},''
  \href{http://dx.doi.org/10.1088/1475-7516/2025/04/012}{{\em JCAP} {\bfseries
  04} (2025) 012}, \href{http://arxiv.org/abs/2404.03000}{{\ttfamily
  arXiv:2404.03000 [astro-ph.CO]}}.

\bibitem{DESI:2024mwx}
{\bfseries DESI} Collaboration, A.~G. Adame {\em et~al.}, ``{DESI 2024 VI:
  cosmological constraints from the measurements of baryon acoustic
  oscillations},'' \href{http://dx.doi.org/10.1088/1475-7516/2025/02/021}{{\em
  JCAP} {\bfseries 02} (2025) 021},
  \href{http://arxiv.org/abs/2404.03002}{{\ttfamily arXiv:2404.03002
  [astro-ph.CO]}}.

\bibitem{Bezrukov:2017dyv}
F.~Bezrukov, M.~Pauly, and J.~Rubio, ``{On the robustness of the primordial
  power spectrum in renormalized Higgs inflation},''
  \href{http://dx.doi.org/10.1088/1475-7516/2018/02/040}{{\em JCAP} {\bfseries
  02} (2018) 040}, \href{http://arxiv.org/abs/1706.05007}{{\ttfamily
  arXiv:1706.05007 [hep-ph]}}.

\bibitem{Barbon:2009ya}
J.~L.~F. Barbon and J.~R. Espinosa, ``{On the Naturalness of Higgs
  Inflation},'' \href{http://dx.doi.org/10.1103/PhysRevD.79.081302}{{\em Phys.
  Rev. D} {\bfseries 79} (2009) 081302},
  \href{http://arxiv.org/abs/0903.0355}{{\ttfamily arXiv:0903.0355 [hep-ph]}}.

\bibitem{Burgess:2010zq}
C.~P. Burgess, H.~M. Lee, and M.~Trott, ``{Comment on Higgs Inflation and
  Naturalness},'' \href{http://dx.doi.org/10.1007/JHEP07(2010)007}{{\em JHEP}
  {\bfseries 07} (2010) 007}, \href{http://arxiv.org/abs/1002.2730}{{\ttfamily
  arXiv:1002.2730 [hep-ph]}}.

\bibitem{Bezrukov:2010jz}
F.~Bezrukov, A.~Magnin, M.~Shaposhnikov, and S.~Sibiryakov, ``{Higgs inflation:
  consistency and generalisations},''
  \href{http://dx.doi.org/10.1007/JHEP01(2011)016}{{\em JHEP} {\bfseries 01}
  (2011) 016}, \href{http://arxiv.org/abs/1008.5157}{{\ttfamily arXiv:1008.5157
  [hep-ph]}}.

\bibitem{Giudice:2010ka}
G.~F. Giudice and H.~M. Lee, ``{Unitarizing Higgs Inflation},''
  \href{http://dx.doi.org/10.1016/j.physletb.2010.10.035}{{\em Phys. Lett. B}
  {\bfseries 694} (2011) 294--300},
  \href{http://arxiv.org/abs/1010.1417}{{\ttfamily arXiv:1010.1417 [hep-ph]}}.

\bibitem{Bauer:2010jg}
F.~Bauer and D.~A. Demir, ``{Higgs-Palatini Inflation and Unitarity},''
  \href{http://dx.doi.org/10.1016/j.physletb.2011.03.042}{{\em Phys. Lett. B}
  {\bfseries 698} (2011) 425--429},
  \href{http://arxiv.org/abs/1012.2900}{{\ttfamily arXiv:1012.2900 [hep-ph]}}.

\bibitem{Rasanen:2017ivk}
S.~Rasanen and P.~Wahlman, ``{Higgs inflation with loop corrections in the
  Palatini formulation},''
  \href{http://dx.doi.org/10.1088/1475-7516/2017/11/047}{{\em JCAP} {\bfseries
  11} (2017) 047}, \href{http://arxiv.org/abs/1709.07853}{{\ttfamily
  arXiv:1709.07853 [astro-ph.CO]}}.

\bibitem{Shaposhnikov:2020fdv}
M.~Shaposhnikov, A.~Shkerin, and S.~Zell, ``{Quantum Effects in Palatini Higgs
  Inflation},'' \href{http://dx.doi.org/10.1088/1475-7516/2020/07/064}{{\em
  JCAP} {\bfseries 07} (2020) 064},
  \href{http://arxiv.org/abs/2002.07105}{{\ttfamily arXiv:2002.07105
  [hep-ph]}}.

\bibitem{Langvik:2020nrs}
M.~Langvik, J.-M. Ojanper{\"a}, S.~Raatikainen, and S.~Rasanen, ``{Higgs
  inflation with the Holst and the Nieh{\textendash}Yan term},''
  \href{http://dx.doi.org/10.1103/PhysRevD.103.083514}{{\em Phys. Rev. D}
  {\bfseries 103} no.~8, (2021) 083514},
  \href{http://arxiv.org/abs/2007.12595}{{\ttfamily arXiv:2007.12595
  [astro-ph.CO]}}.

\bibitem{Shaposhnikov:2025eyp}
M.~Shaposhnikov, ``Higgs particle and cosmology,'' in {\em Encyclopedia of
  Particle Physics}.
\newblock Elsevier, 2025.
\newblock \url{https://doi.org/10.1016/b978-0-443-26598-3.00014-6}.

\bibitem{Bezrukov:2014ipa}
F.~Bezrukov, J.~Rubio, and M.~Shaposhnikov, ``{Living beyond the edge: Higgs
  inflation and vacuum metastability},''
  \href{http://dx.doi.org/10.1103/PhysRevD.92.083512}{{\em Phys. Rev. D}
  {\bfseries 92} no.~8, (2015) 083512},
  \href{http://arxiv.org/abs/1412.3811}{{\ttfamily arXiv:1412.3811 [hep-ph]}}.

\bibitem{Palatini:1919ffw}
A.~Palatini, ``{Deduzione invariantiva delle equazioni gravitazionali dal
  principio di Hamilton},'' \href{http://dx.doi.org/10.1007/BF03014670}{{\em
  Rend. Circ. Mat. Palermo} {\bfseries 43} no.~1, (1919) 203--212}.

\bibitem{Hehl:1976kj}
F.~W. Hehl, P.~Von Der~Heyde, G.~D. Kerlick, and J.~M. Nester, ``{General
  Relativity with Spin and Torsion: Foundations and Prospects},''
  \href{http://dx.doi.org/10.1103/RevModPhys.48.393}{{\em Rev. Mod. Phys.}
  {\bfseries 48} (1976) 393--416}.

\bibitem{Cartan:1922}
{\'E}.~Cartan, ``{Sur une g{\'e}n{\'e}ralisation de la notion de courbure de
  Riemann et les espaces {\`a} torsion},'' {\em Comptes Rendus, Ac. Sc. Paris}
  {\bfseries 174} (1922) 593--595.

\bibitem{Cartan:1923}
{\'E}.~Cartan, ``{Sur les vari{\'e}t{\'e}s {\`a} connexion affine et la
  th{\'e}orie de la relativit{\'e} g{\'e}n{\'e}ralis{\'e}e (premi{\`e}re
  partie)},'' in {\em Annales scientifiques de l'{\'E}cole normale
  sup{\'e}rieure}, vol.~40, pp.~325--412.
\newblock Gauthier-Villars, 1923.

\bibitem{Cartan:1924}
{\'E}.~Cartan, ``{Sur les vari{\'e}t{\'e}s {\`a} connexion affine, et la
  th{\'e}orie de la relativit{\'e} g{\'e}n{\'e}ralis{\'e}e (premi{\`e}re
  partie)(suite)},'' in {\em Annales scientifiques de l'{\'E}cole Normale
  Sup{\'e}rieure}, vol.~41, pp.~1--25.
\newblock Gauthier-Villars, 1924.

\bibitem{Cartan:1925}
{\'E}.~Cartan, ``{Sur les vari{\'e}t{\'e}s {\`a} connexion affine, et la
  th{\'e}orie de la relativit{\'e} g{\'e}n{\'e}ralis{\'e}e (deuxi{\`e}me
  partie)},'' in {\em Annales scientifiques de l'{\'E}cole normale
  sup{\'e}rieure}, vol.~42, pp.~17--88.
\newblock Gauthier-Villars, 1925.

\bibitem{Utiyama:1956sy}
R.~Utiyama, ``{Invariant theoretical interpretation of interaction},''
  \href{http://dx.doi.org/10.1103/PhysRev.101.1597}{{\em Phys. Rev.} {\bfseries
  101} (1956) 1597--1607}.

\bibitem{Kibble:1961ba}
T.~W.~B. Kibble, ``{Lorentz invariance and the gravitational field},''
  \href{http://dx.doi.org/10.1063/1.1703702}{{\em J. Math. Phys.} {\bfseries 2}
  (1961) 212--221}.

\bibitem{Sciama:1962}
D.~W. Sciama, ``On the analogy between charge and spin in general relativity,''
  in {\em Recent developments in general relativity}, p.~415.
\newblock Pergamon Press, Oxford, 1962.

\bibitem{Shaposhnikov:2020gts}
M.~Shaposhnikov, A.~Shkerin, I.~Timiryasov, and S.~Zell, ``{Higgs inflation in
  Einstein-Cartan gravity},''
  \href{http://dx.doi.org/10.1088/1475-7516/2021/10/E01}{{\em JCAP} {\bfseries
  02} (2021) 008}, \href{http://arxiv.org/abs/2007.14978}{{\ttfamily
  arXiv:2007.14978 [hep-ph]}}. [Erratum: JCAP 10, E01 (2021)].

\bibitem{Karananas:2021zkl}
G.~K. Karananas, M.~Shaposhnikov, A.~Shkerin, and S.~Zell, ``{Matter matters in
  Einstein-Cartan gravity},''
  \href{http://dx.doi.org/10.1103/PhysRevD.104.064036}{{\em Phys. Rev. D}
  {\bfseries 104} no.~6, (2021) 064036},
  \href{http://arxiv.org/abs/2106.13811}{{\ttfamily arXiv:2106.13811
  [hep-th]}}.

\bibitem{Bauer:2008zj}
F.~Bauer and D.~A. Demir, ``{Inflation with Non-Minimal Coupling: Metric versus
  Palatini Formulations},''
  \href{http://dx.doi.org/10.1016/j.physletb.2008.06.014}{{\em Phys. Lett. B}
  {\bfseries 665} (2008) 222--226},
  \href{http://arxiv.org/abs/0803.2664}{{\ttfamily arXiv:0803.2664 [hep-ph]}}.

\bibitem{Obukhov:1982zn}
Y.~N. Obukhov, ``{Conformal invariance and space-time torsion},''
  \href{http://dx.doi.org/10.1016/0375-9601(82)90037-8}{{\em Phys. Lett. A}
  {\bfseries 90} (1982) 13--16}.

\bibitem{Dereli:1982xb}
T.~Dereli and R.~W. Tucker, ``{Weyl scalings and spinor matter interactions in
  scalar - tensor theories of gravitation},''
  \href{http://dx.doi.org/10.1016/0370-2693(82)91237-0}{{\em Phys. Lett. B}
  {\bfseries 110} (1982) 206--210}.

\bibitem{Karananas:2024xja}
G.~K. Karananas, M.~Shaposhnikov, and S.~Zell, ``{Weyl-invariant
  Einstein-Cartan gravity: unifying the strong CP and hierarchy puzzles},''
  \href{http://dx.doi.org/10.1007/JHEP11(2024)146}{{\em JHEP} {\bfseries 11}
  (2024) 146}, \href{http://arxiv.org/abs/2406.11956}{{\ttfamily
  arXiv:2406.11956 [hep-th]}}.

\bibitem{Atkins:2012yn}
M.~Atkins and X.~Calmet, ``{Bounds on the Nonminimal Coupling of the Higgs
  Boson to Gravity},''
  \href{http://dx.doi.org/10.1103/PhysRevLett.110.051301}{{\em Phys. Rev.
  Lett.} {\bfseries 110} no.~5, (2013) 051301},
  \href{http://arxiv.org/abs/1211.0281}{{\ttfamily arXiv:1211.0281 [hep-ph]}}.

\bibitem{Englert:1976ep}
F.~Englert, C.~Truffin, and R.~Gastmans, ``{Conformal Invariance in Quantum
  Gravity},'' \href{http://dx.doi.org/10.1016/0550-3213(76)90406-5}{{\em Nucl.
  Phys. B} {\bfseries 117} (1976) 407--432}.

\bibitem{Wetterich:1987fm}
C.~Wetterich, ``{Cosmology and the Fate of Dilatation Symmetry},''
  \href{http://dx.doi.org/10.1016/0550-3213(88)90193-9}{{\em Nucl. Phys. B}
  {\bfseries 302} (1988) 668--696},
  \href{http://arxiv.org/abs/1711.03844}{{\ttfamily arXiv:1711.03844
  [hep-th]}}.

\bibitem{Shaposhnikov:2008xi}
M.~Shaposhnikov and D.~Zenhausern, ``{Quantum scale invariance, cosmological
  constant and hierarchy problem},''
  \href{http://dx.doi.org/10.1016/j.physletb.2008.11.041}{{\em Phys. Lett. B}
  {\bfseries 671} (2009) 162--166},
  \href{http://arxiv.org/abs/0809.3406}{{\ttfamily arXiv:0809.3406 [hep-th]}}.

\bibitem{Ghilencea:2015mza}
D.~M. Ghilencea, ``{Manifestly scale-invariant regularization and quantum
  effective operators},''
  \href{http://dx.doi.org/10.1103/PhysRevD.93.105006}{{\em Phys. Rev. D}
  {\bfseries 93} no.~10, (2016) 105006},
  \href{http://arxiv.org/abs/1508.00595}{{\ttfamily arXiv:1508.00595
  [hep-ph]}}.

\bibitem{Ghilencea:2016dsl}
D.~M. Ghilencea, Z.~Lalak, and P.~Olszewski, ``{Standard Model with
  spontaneously broken quantum scale invariance},''
  \href{http://dx.doi.org/10.1103/PhysRevD.96.055034}{{\em Phys. Rev. D}
  {\bfseries 96} no.~5, (2017) 055034},
  \href{http://arxiv.org/abs/1612.09120}{{\ttfamily arXiv:1612.09120
  [hep-ph]}}.

\bibitem{Ghilencea:2017yqv}
D.~M. Ghilencea, ``{Quantum implications of a scale invariant
  regularization},'' \href{http://dx.doi.org/10.1103/PhysRevD.97.075015}{{\em
  Phys. Rev. D} {\bfseries 97} no.~7, (2018) 075015},
  \href{http://arxiv.org/abs/1712.06024}{{\ttfamily arXiv:1712.06024
  [hep-th]}}.

\bibitem{Ghilencea:2016ckm}
D.~M. Ghilencea, Z.~Lalak, and P.~Olszewski, ``{Two-loop scale-invariant scalar
  potential and quantum effective operators},''
  \href{http://dx.doi.org/10.1140/epjc/s10052-016-4475-0}{{\em Eur. Phys. J. C}
  {\bfseries 76} no.~12, (2016) 656},
  \href{http://arxiv.org/abs/1608.05336}{{\ttfamily arXiv:1608.05336
  [hep-th]}}.

\bibitem{Ferreira:2018itt}
P.~G. Ferreira, C.~T. Hill, and G.~G. Ross, ``{Inertial Spontaneous Symmetry
  Breaking and Quantum Scale Invariance},''
  \href{http://dx.doi.org/10.1103/PhysRevD.98.116012}{{\em Phys. Rev. D}
  {\bfseries 98} no.~11, (2018) 116012},
  \href{http://arxiv.org/abs/1801.07676}{{\ttfamily arXiv:1801.07676
  [hep-th]}}.

\bibitem{Ghilencea:2024usf}
D.~M. Ghilencea, ``{Quantum gravity from Weyl conformal geometry},''
  \href{http://dx.doi.org/10.1140/epjc/s10052-025-14489-z}{{\em Eur. Phys. J.
  C} {\bfseries 85} no.~7, (2025) 815},
  \href{http://arxiv.org/abs/2408.07160}{{\ttfamily arXiv:2408.07160
  [hep-th]}}.

\bibitem{Shaposhnikov:2025znm}
M.~Shaposhnikov, {\em {Progress in Einstein-Cartan gravity}}.
\newblock 6, 2025.
\newblock \href{http://arxiv.org/abs/2506.11847}{{\ttfamily arXiv:2506.11847
  [hep-th]}}.

\bibitem{Shaposhnikov:2018xkv}
M.~Shaposhnikov and A.~Shkerin, ``{Conformal symmetry: towards the link between
  the Fermi and the Planck scales},''
  \href{http://dx.doi.org/10.1016/j.physletb.2018.06.068}{{\em Phys. Lett. B}
  {\bfseries 783} (2018) 253--262},
  \href{http://arxiv.org/abs/1803.08907}{{\ttfamily arXiv:1803.08907
  [hep-th]}}.

\bibitem{Shaposhnikov:2018jag}
M.~Shaposhnikov and A.~Shkerin, ``{Gravity, Scale Invariance and the Hierarchy
  Problem},'' \href{http://dx.doi.org/10.1007/JHEP10(2018)024}{{\em JHEP}
  {\bfseries 10} (2018) 024}, \href{http://arxiv.org/abs/1804.06376}{{\ttfamily
  arXiv:1804.06376 [hep-th]}}.

\bibitem{Shaposhnikov:2020geh}
M.~Shaposhnikov, A.~Shkerin, and S.~Zell, ``{Standard Model Meets Gravity:
  Electroweak Symmetry Breaking and Inflation},''
  \href{http://dx.doi.org/10.1103/PhysRevD.103.033006}{{\em Phys. Rev. D}
  {\bfseries 103} no.~3, (2021) 033006},
  \href{http://arxiv.org/abs/2001.09088}{{\ttfamily arXiv:2001.09088
  [hep-th]}}.

\bibitem{Karananas:2025ews}
G.~K. Karananas, M.~Shaposhnikov, and S.~Zell, ``{Gravitational Origin of the
  QCD Axion},'' \href{http://dx.doi.org/10.1103/d7mp-sjvc}{{\em Phys. Rev.
  Lett.} {\bfseries 135} no.~24, (2025) 241001},
  \href{http://arxiv.org/abs/2506.11836}{{\ttfamily arXiv:2506.11836
  [hep-th]}}.

\bibitem{Karananas:2025uhy}
G.~K. Karananas, M.~Shaposhnikov, and S.~Zell, ``{A non-compact QCD axion},''
  \href{http://arxiv.org/abs/2512.20290}{{\ttfamily arXiv:2512.20290
  [hep-ph]}}.

\bibitem{Peccei:1977hh}
R.~D. Peccei and H.~R. Quinn, ``{CP Conservation in the Presence of
  Instantons},'' \href{http://dx.doi.org/10.1103/PhysRevLett.38.1440}{{\em
  Phys. Rev. Lett.} {\bfseries 38} (1977) 1440--1443}.

\bibitem{Karananas:2025qsm}
G.~K. Karananas and M.~Shaposhnikov, ``{Higgs inflation in Weyl-invariant
  Einstein-Cartan gravity},''
  \href{http://dx.doi.org/10.1016/j.physletb.2026.140376}{{\em Phys. Lett. B}
  {\bfseries 876} (2026) 140376},
  \href{http://arxiv.org/abs/2511.04732}{{\ttfamily arXiv:2511.04732
  [hep-ph]}}.

\end{thebibliography}\endgroup

\end{document}